\documentclass{aa}  
\usepackage{graphicx}
\usepackage{subcaption}
\usepackage[colorlinks=true, citecolor=blue, linkcolor=blue, urlcolor=black]{hyperref}
\usepackage[varg]{txfonts}

\begin{document}

   \title{LOFAR properties of SILVERRUSH Ly$\alpha$ emitter candidates in the ELAIS-N1 field}
   \author{A. J. Gloudemans \inst{\ref{inst1}}
   \and K. J. Duncan \inst{\ref{inst1}, \ref{inst2}} 
   \and R. Kondapally \inst{\ref{inst2}} 
   \and J. Sabater \inst{\ref{inst2}} 
   \and R. K. Cochrane \inst{\ref{inst3}} 
   \and H. J. A. R\"{o}ttgering \inst{\ref{inst1}}   
   \and P. N. Best \inst{\ref{inst2}} 
   \and M. Bonato \inst{\ref{inst4}, \ref{inst4alma}, \ref{inst5}} 
   \and M. Bondi \inst{\ref{inst4}}
   \and K. Malek \inst{\ref{inst_poland}, \ref{inst_marseille}} 
   \and I. McCheyne \inst{\ref{inst_sussex}} 
   \and D. J. B. Smith \inst{\ref{inst_Hertfordshire}} 
   \and I. Prandoni \inst{\ref{inst4}} 
   \and L. Wang \inst{\ref{inst6}, \ref{inst7}}}

   \institute{Leiden Observatory, Leiden University, PO Box 9513, 2300 RA Leiden, The Netherlands \\ e-mail: gloudemans@strw.leidenuniv.nl\label{inst1} 
   \and SUPA, Institute for Astronomy, Royal Observatory, Blackford Hill, Edinburgh, EH9 3HJ, UK\label{inst2} \and Center for Astrophysics, Harvard \& Smithsonian, 60 Garden St, Cambridge, MA, 02138, USA\label{inst3} \and INAF-Istituto di Radioastronomia, Via Gobetti 101, I-40129, Bologna, Italy \label{inst4} 
   \and Italian ALMA Regional Centre, Via Gobetti 101, I-40129, Bologna, Italy\label{inst4alma}
   \and INAF-Osservatorio Astronomico di Padova, Vicolo dell'Osservatorio 5, I-35122, Padova, Italy\label{inst5} 
   \and National Centre for Nuclear Research, ul. Pasteura 7, 02-093 Warszawa, Poland\label{inst_poland}
   \and Aix Marseille Univ. CNRS, CNES, LAM, Marseille, France\label{inst_marseille}
   \and Astronomy Centre, Dept. of Physics \& Astronomy, University of Sussex, Brighton BN1 9QH, UK\label{inst_sussex}
   \and Centre for Astrophysics Research, School of Physics, Astronomy and Mathematics, University of Hertfordshire, College Lane, Hatfield AL10 9AB, UK\label{inst_Hertfordshire}
   \and SRON Netherlands Institute for Space Research, Landleven 12, 9747 AD, Groningen, The Netherlands\label{inst6} 
   \and Kapteyn Astronomical Institute, University of Groningen, Postbus 800, 9700 AV Groningen, the Netherlands\label{inst7}}

   \date{Received: 01 July 2020 / Accepted: 16 October 2020}


    \abstract{Lyman alpha emitters  (LAEs) in the Epoch of Reionization (EoR) offer valuable probes of both early galaxy evolution and the process of reionization itself; however, the exact evolution of their abundance and the nature of their emission remain open questions.
    We combine samples of 229 and 349 LAE candidates at $z=5.7$ and $z=6.6,$ respectively, from the SILVERRUSH narrowband survey with deep Low Frequency Array (LOFAR) radio continuum observations in the European Large Area Infrared Space Observatory Survey-North 1 (ELAIS-N1) field to search for radio galaxies in the EoR and study the low-frequency radio properties of $z\gtrsim5.7$ LAE emitters. Our LOFAR observations reach an unprecedented noise level of $\sim20\,\mu$Jy beam$^{-1}$ at 150MHz, and we detect five candidate LAEs at $>5\sigma$ significance. Based on detailed spectral energy distribution (SED) modelling of independent multi-wavelength observations in the field, we conclude that these sources are likely [O\textsc{ii}] emitters at $z=1.47$, yielding no reliable $z\gtrsim5.7$ radio galaxy candidates. We examine the 111 $z=5.7$ and $z=6.6$ LAE candidates from our panchromatic photometry catalogue  that are undetected by LOFAR, finding contamination rates of 81-92\% for the $z=5.7$ and $z=6.6$ subset of the LAE candidate samples. This subset of the full sample is biased towards brighter magnitudes and redder near-infrared colours. The contamination rates of the full sample will therefore likely be lower than the reported values. Contamination of these optically bright LAE samples by likely [O\textsc{ii}] emitters is lowered significantly through constraints on the near-infrared colours, highlighting the need for infrared observations to robustly identify bright LAEs in narrowband surveys. Finally, the stacking of radio continuum observations for the robust LAE samples yields 2$\sigma$ upper limits on radio luminosity of 8.2$\times$10$^{23}$ and 8.7$\times$10$^{23}$ W Hz$^{-1}$ at $z=5.7$ and $6.6$, respectively, corresponding to limits on their median star-formation rates of $<$53 and $<$56 M$_{\odot}$ yr$^{-1}$.}


\defcitealias{Duncan2020}{D20}

   \maketitle


\section{Introduction}

Luminous Lyman alpha emitters (LAEs) at high redshift offer valuable probes of galaxy evolution and cosmology in the early Universe.
The existence of young galaxies in the early phase of their evolution with strong Lyman alpha (Ly$\alpha)$ emission (1216 $\r{A}$) was first hypothesised by \cite{Patridge1967ApJ...147..868P} and first observed by \cite{Hu1996Natur.382..231H} and \cite{Pascarelle1996Natur.383...45P}. Dedicated narrowband, spectroscopic, and integral field unit surveys have since discovered thousands of LAEs at redshifts 2 < z < 7, opening up a new way of studying the high-redshift Universe (e.g. \citealt{cowie1998high, Rhoads2000ApJ...545L..85R, Ouchi2003ApJ...582...60O, Ciardullo_2011, Shibuya2012ApJ...752..114S, Konno2014ApJ...797...16K, Sobral2018MNRAS.476.4725S}). Previous studies have identified two populations of LAEs: blue and faint LAEs with low masses and metallicities (e.g. \citealt{Bacon2015A&A...575A..75B, Sobral2015ApJ...808..139S, Sobral2019MNRAS.482.2422S, Nakajima2016ApJ...831L...9N, Ono2010MNRAS.402.1580O}) and red, massive, and luminous LAEs (e.g. \citealt{chapman2005ApJ...622..772C, Sandberg2015A&A...580A..91S, Matthee2016MNRAS.458..449M}) often observed to host an active galactic nucleus (AGN) (e.g. \citealt{Ouchi2008ApJS..176..301O, Konno2016ApJ...823...20K, Sobral2017MNRAS.466.1242S}). 

Lyman alpha emitters have been widely used to study luminosity functions (LFs) and clustering properties of galaxies in the early Universe (e.g. \citealt{Ouchi2003ApJ...582...60O, Shimasaku2004ApJ...605L..93S, Ouchi2010ApJ...723..869O, Kusakabe2018PASJ...70....4K, khostovan2019MNRAS.489..555K}). Recent studies, for example \cite{khostovan2019MNRAS.489..555K},
suggest that LAEs are progenitors of a wide range of galaxy types, where the brightest LAEs are located in the most massive halos and are highly clustered. The LAEs that have been detected at radio wavelengths have been found to have steep radio spectral indices \citep{calhau2020MNRAS.tmp..448C}.

Furthermore, studies have shown that the AGN fraction rises with Ly$\alpha$ luminosity, leading to the AGN dominating (over the star-forming population) at the bright end of the Ly$\alpha$ LF (e.g. \citealt{Sobral2018MNRAS.477.2817S, Sobral2018MNRAS.476.4725S, Matthee2017MNRAS.472..772M, Wold2014ApJ...783..119W, Wold2017ApJ...848..108W, calhau2020MNRAS.tmp..448C}). As suggested by \cite{Sobral2018MNRAS.477.2817S} and demonstrated by \cite{calhau2020MNRAS.tmp..448C}, the AGN fraction of LAEs declines towards higher redshifts at a fixed Ly$\alpha$ luminosity. However, these conclusions are limited by the number of detected LAEs at high redshift, and larger samples are required to confirm this claim. Constraining the AGN fraction in the Ly$\alpha$ LF is vital for our understanding of early super massive black hole (SMBH) formation (e.g. \citealt{calhau2020MNRAS.tmp..448C}), but it is also critical for understanding the source of ionising photons of LAEs if they are to be used as probes of reionization (e.g. \citealt{Santos2016MNRAS.463.1678S, Matthee2015MNRAS.451..400M}).

The Low Frequency Array (LOFAR) Two-metre Sky Survey  (LoTSS; \citealt{Shimwell2017A&A...598A.104S,Shimwell2019A&A...622A...1S}) is entering a new regime of deep, low-frequency surveys by pushing noise levels to below 100 $\mu$Jy beam$^{-1}$ at 150 MHz across the entire northern sky and complementing this with targeted, deeper observations in the degree-scale northern deep fields. The LoTSS Deep Fields first data release \citep{Tasse2020, Sabater2020, Kondapally2020, Duncan2020} covers a total area of 25 deg$^2$ in the Lockman Hole, European Large Area Infrared Space Observatory Survey-North 1 (hereafter ELAIS-N1), and Bo\"otes fields, at 6" resolution, reaching an rms depth of S$_{150\rm{MHz}}$ $\sim$ 20 $\mu$Jy beam$^{-1}$ in the deepest field, ELAIS-N1.

One of the LoTSS target fields, ELAIS-N1 (RA=242.75, Dec= 54.95 degrees), has also been the target field for a dedicated LAE survey named the Systematic Identification of LAEs for Visible Exploration and Reionization Research Using Subaru HSC (SILVERRUSH; \citealt{Ouchi2018}). The SILVERRUSH program makes use of the narrowband observations of the Hyper Suprime-Cam Subaru Strategic Program (HSC-SSP; \citealt{Aihara2018PASJ...70S...4A}). Three major scientific goals of the SILVERRUSH program are studying LAE properties at high redshift, using LAEs to probe the low-mass young galaxy population, and studying the Ly$\alpha$ LF towards the Epoch of Reionization (EoR) \citep{Ouchi2018}. The SILVERRUSH program identified $\sim$2000 LAE candidates at $z =$ 5.7 and $z =$ 6.6 across four different fields, $\sim$600 of which are located in ELAIS-N1 \citep{Shibuya2018}. The overlapping LoTSS and SILVERRUSH observations of ELAIS-N1 open up a new opportunity to study the as yet unexplored low-frequency (150 MHz) radio properties of LAEs. Furthermore, the identification of LAE radio AGN would allow us to study the AGN fraction as a function of Ly$\alpha$ luminosity and follow up on the claim that the LAE AGN fraction declines towards higher redshifts. Finally, according to model predictions by \cite{Saxena2017MNRAS.469.4083S}, more than ten radio-loud AGN at $z > 6$ are expected to be detected in all the LOFAR Deep Fields combined with the current sensitivity. Currently, the most distant radio galaxy has a redshift of $z= 5.72$ \citep{Saxena2018MNRAS.480.2733S}. The SILVERRUSH catalogue  is a potentially excellent source for finding radio galaxies at $z > 6$, even though only a small fraction of the volume is probed due to the detection range of the narrowband filters. The detection of such a high-redshift radio galaxy would not only be a substantial step forwards in the study of the formation and evolution of massive galaxies into the EoR, but it would also be the most distant radio galaxy discovered to date. 

This paper is structured as follows. In Sect. \ref{sec:Data}, we describe the details of the SILVERRUSH program, the LoTSS Deep Fields data used, and the LAE sample selection. In Sect. \ref{sec:radio_silverrush}, we present the characteristics of the radio-detected SILVERRUSH population. Subsequently, in Sect. \ref{sec:radio_wider_pop}, we analyse the optical and infrared (IR) properties and multi-wavelength stacks of the wider LAE population. In Sect. \ref{sec:discussion}, we discuss the LOFAR detection rate and LAE selection. Finally, in Sect. \ref{sec:summary}, we summarise our findings.

In this work, a flat lambda cold dark matter ($\Lambda$-CDM) cosmology is assumed using H$_{0}$= 70 km s$^{-1}$ Mpc$^{-1}$, $\Omega_{M}$ = 0.3, and $\Omega_{\Lambda}$ = 0.7. Furthermore, all magnitudes presented are given in the AB system \citep{Oke1983ApJ...266..713O}.

\section{Data and sample selection}
\label{sec:Data}

\subsection{SILVERRUSH program} 

To identify LAEs, the SILVERRUSH survey uses four narrowband (hereafter NB) filters, which are mounted on the Hyper Suprime-Cam (HSC). These filters are NB387, NB816, NB921, and NB101, corresponding to central wavelengths ($\lambda_{\text{cen}}$) of 3858, 8169, 9204, and 10092 $\r{A}$\footnote{\url{http://svo2.cab.inta-csic.es/theory/fps/index.php?mode=browse&gname=Subaru&gname2=HSC}}, which allow for the identification of LAEs at redshifts of $z=2.17\pm0.02$, 5.72$\pm$0.05, 6.57$\pm$0.05, and 7.30$\pm$0.04, respectively, in addition to rest frame optical emission line galaxies at lower redshifts \citep[see][]{Aihara2018PASJ...70S...4A}. Using NB816 and NB921 imaging data from the HSC-SSP survey, the SILVERRUSH program identified $\sim$2000 LAE candidates at $z=5.7$ and $z=6.6$ in five fields, with a total area of 14 and 21 deg$^2$, respectively \citep{Shibuya2018}. These resulting LAE catalogues are publicly available\footnote{\url{http://cos.icrr.u-tokyo.ac.jp/rush.html}} and will be referred to as the NB816 and NB921 LAE catalogues in this work. The public data release does not include the results of LAE studies based on NB387 and NB101 observations. The detailed SILVERRUSH program strategy is described by \cite{Ouchi2018}, and the first SILVERRUSH catalogues and properties are presented by \cite{Shibuya2018}. Other main scientific results following from SILVERRUSH, such as clustering properties and Ly$\alpha$ LFs, have been published in \cite{Shibuya2018_SRIII}, \cite{Konno2018_SR4}, \cite{Harikane2018_SR5}, \cite{Inoue2018_SR6}, \cite{Higuchi2019_SR7}, \cite{Harikane2019_SR8}, and \cite{Kakuma2019_SR9}.
In the following sections, we focus on the data and sample selection in the ELAIS-N1 field. The data obtained by the HSC-SSP survey are outlined in Sect. \ref{subsubsec:hsc_ssp}, and the SILVERRUSH procedure for LAE selection from this survey is summarised in Sect. \ref{subsubsection:LAE_selection}.

\subsubsection{HSC-SSP survey}
\label{subsubsec:hsc_ssp}
\begin{figure}
\centering
   \includegraphics[width=\columnwidth, trim={0.0cm 0cm 0cm 0.0cm}, clip]{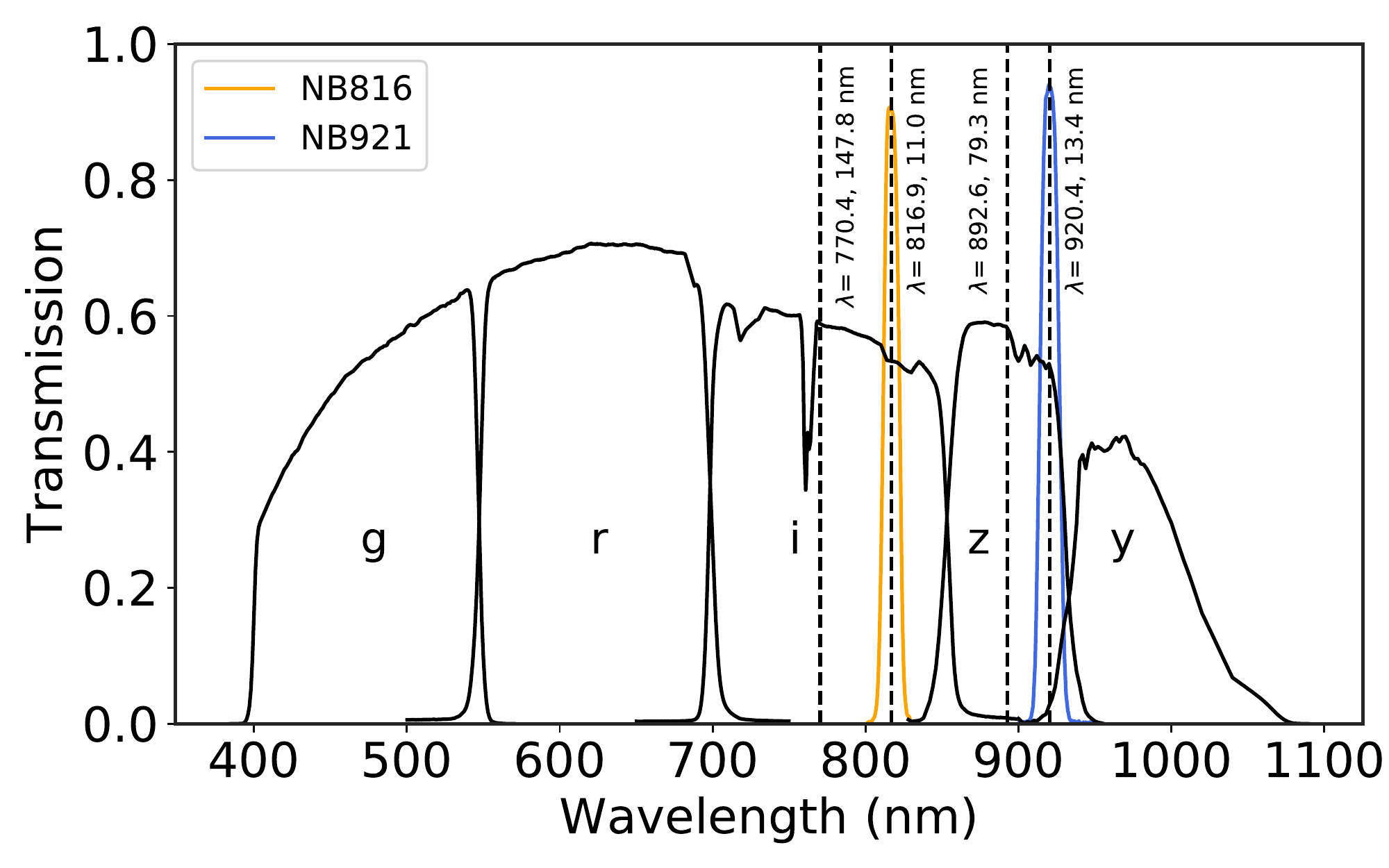}
     \caption{Filter transmission curves of HSC BB (\textit{grizy}) and NB (NB816 \& NB921) filters. The central wavelength ($\lambda_{\text{cen}}$) and full width at  half maximum (in nm) are indicated for the \textit{i}, \textit{z}, NB816, and NB921 filters. The filter transmission curves and $\lambda_{\text{cen}}$ values have been obtained from the SVO Filter Profile Service$^1$.} 
     \label{fig:hsc_filters}
\end{figure}

The HSC filter transmission curves of broadbands (BBs) \textit{g, r, i, z}, and \textit{y} and narrowbands NB816 and NB921 are shown in Fig. \ref{fig:hsc_filters}, including the central wavelengths $\lambda_{\text{cen}}$ and full width half     maxima (FWHM) for the \textit{i} and \textit{z} bands and NB filters. The NB816 and NB921 LAE catalogues were created using HSC-SSP S16A data taken using these five BB and NB filters \citep{Ouchi2018}. Deep observations of ELAIS-N1 were conducted as part of the HSC-SSP survey, reaching 5$\sigma$ limiting magnitudes of 25.3 for the NBs and $\sim$24-26 for the BBs \citep{Shibuya2018}. The image reduction was performed using the HSC pipeline \citep{bosch2018hyper}. The source detection and photometric measurements are obtained using a `forced' and `unforced' method. In the unforced method, the coordinates, shape, and flux of each source are determined individually in each band, whereas in the forced method the coordinates and shape of the sources are fixed in a reference band and applied to all other bands to determine the flux. In this work, the photometry from the both forced and unforced methods is used to ensure the largest possible sample. Further details on the HSC-SSP survey is available in \cite{Aihara2018PASJ...70S...4A} and \cite{Shibuya2018}.

\subsubsection{LAE selection}
\label{subsubsection:LAE_selection}

The LAEs in the SILVERRUSH program were selected using colour selection criteria (presented in \citealt{Shibuya2018} and based on \citealt{Ouchi2008ApJS..176..301O, Ouchi2010ApJ...723..869O}), which ensure a non-detection in bands blue-wards of the Lyman break at a certain redshift and significant detected flux excess in the NBs. These criteria are given by:
\begin{align}
\label{eq:idrop}
    i - NB816 &\geq 1.2 \nonumber \\
    g &> g_{3\sigma} \\
    (r \leq r_{3\sigma} \ \text{and} \ r-i \geq 1.0) & \ \text{or} \ (r > r_{3\sigma}) \nonumber
\end{align} 
\noindent and 
\begin{align}
\label{eq:zdrop}
    z - NB921 &\geq 1.0 \nonumber \\
    g > g_{3\sigma} \ & \text{and} \ r > r_{3\sigma}\\
    (z \leq z_{3\sigma} \ \text{and} \ i-z \geq 1.3) & \ \text{or} \ (z > z_{3\sigma}) \nonumber
\end{align} 
\noindent for the $z=5.7$ and $z=6.6$ LAEs respectively, where \textit{g}$_{3\sigma}$, \textit{r}$_{3\sigma}$, and \textit{z}$_{3\sigma}$ are the $3\sigma$ limiting magnitudes in the \textit{g, r,} and \textit{z }bands. In the forced catalogue, a stricter colour selection criterion of z$-$NB921 > 1.8 is used, and the limiting magnitudes are lowered to 2$\sigma$ in the \textit{g} and \textit{r} bands. These stricter colour criteria are also used in \cite{Konno2018_SR4} for studying the Ly$\alpha$ LF. Besides the colour selection, several parameters and flags were used to denote imaging problems or artefacts, such as sources containing saturated pixels and images with short exposure times (see \citealt{Shibuya2018,Aihara2018PASJ...70S...8A}). Further details on the SILVERRUSH LAE selection algorithm are presented in \cite{Shibuya2018}.

The final SILVERRUSH samples in ELAIS-N1 contain 229 and 349 sources at $z=5.7$ and 6.6, respectively, in a 6 deg$^2$ area \citep{Konno2018_SR4}. The reliability of their LAE selection in all fields was checked using spectroscopic observations of a sub-sample of the selected LAEs. In total, redshifts of 96 of the HSC LAEs were spectroscopically confirmed, and they yielded a contamination rate of 0-30\%, depending on the magnitude ranges \citep{Shibuya2018_SRIII}.

\subsection{LOFAR radio observations of ELAIS-N1}

LoTSS aims to cover the whole Northern Hemisphere, reaching $\leq$100 $\mu$Jy beam$^{-1}$ noise levels; a part of the data was released in the first data release (LoTSS-DR1; \citealt{Shimwell2019A&A...622A...1S}). LoTSS is being complemented by a series of LoTSS Deep Fields, which aim to ultimately cover a sky area of $\sim$50 deg$^2$ down to noise levels of $\sim$10 $\mu$Jy beam$^{-1}$, probing the fainter and higher redshift radio population. The LoTSS Deep Fields DR1 targets the Lockman Hole, Bo\"otes, and ELAIS-N1 fields, which are by design at declinations optimal for LOFAR observing sensitivity. These fields have already been extensively observed across optical and IR wavelengths, enabling the determination of photometric redshifts and physical galaxy properties. New imaging and calibration algorithms have been developed by \cite{Tasse2020} to enable the construction of thermal noise-limited images, and the LoTSS Deep Fields radio images reach $\sim$20 $\mu$Jy beam$^{-1}$ rms sensitivities in ELAIS-N1 \citep{Sabater2020}. Details on the ELAIS-N1 radio imaging and catalogue are presented in \cite{Sabater2020}.

\subsection{Deep optical-IR photometry in ELAIS-N1}
\label{subsec:lofar_optical_phot}

\begin{table*}
\caption{Overview of the multi-wavelength observations in ELAIS-N1 used for this work. The area covered by each survey and typical depths (3$\sigma$) per survey, except for SPIRE, PACS, and MIPS, are adopted from \cite{Kondapally2020}; the detailed 3$\sigma$ depths per filter are available in that paper. The depths for SPIRE, PACS, and MIPS are given by flux limits in mJy. Details on the optical Panoramic Survey Telescope and Rapid Response System (PanSTARRS) Medium Deep Survey, the HSC-SSP survey, and the Canada France Hawaii Telescope (CFHT) MegaCam SpARCS  are available in \cite{Chambers2016arXiv161205560C}, \cite{Aihara2018PASJ...70S...8A}, and \cite{Wilson2009ApJ...698.1943W}, respectively. A detailed description of the UKIRT Infrared Deep Sky Survey - Deep Extragalactic Survey (UKIDSS-DXS) conducted with the Wide Field Camera (WFCAM) on the United Kingdom Infrared Telescope (UKIRT) can be found in \cite{Lawrence2007MNRAS.379.1599L}. The Herschel Multi-tiered Extragalactic Survey (HerMES) is described in \cite{Oliver2012MNRAS.424.1614O}, and the Spitzer Wide-Area Infrared Extragalactic Survey (SWIRE) and Spitzer Extragalactic Representative Volume Survey (SERVS)  are described in \cite{Lonsdale2003PASP..115..897L} and \cite{Mauduit2012PASP..124..714M}, respectively.}
\label{tab:observations}      
\centering
\resizebox{\textwidth}{!}{
\begin{tabular}{c c c c c c}  
\hline\hline
Telescope & Instrument & Survey &  Wavelength cov & Area (deg$^2$) &  Depth \\ 
\hline
LOFAR & & LoTSS Deep Fields & 150 MHz & $\sim$25 & $\sim$20 $\mu$Jy beam$^{-1}$ \\ 
Herschel & SPIRE & HerMES & 250, 350, 500 $\mu$m & 7.16 &  4 mJy\\ 
 & PACS & HerMES & 110, 170 $\mu$m & 7.13 & 12.5/17.5 mJy\\
 Spitzer & MIPS & SWIRE & 24 $\mu$m & 7.16  &  20 mJy\\ 
 & IRAC & SWIRE & 3.6, 4.5, 5.8, 8.0 $\mu$m & 9.32 & $\sim$22.2 mag\\ 
  & & SERVS & 3.6, 4.5 $\mu$m & 2.39 & 24.1 mag\\ 
UKIRT & WFCAM & UKIDSS-DXS  & J and K & 8.87 & $\sim$23.0 mag\\ 
Haleakala Observatory & PanSTARRS & Medium Deep Survey & grizy & 8.05 & $\sim$24.7 mag\\ 
Subaru Telescope & HSC & SSP survey & grizy, NB816, NB921 & 7.70 &  $\sim$24.5 mag\\ 
CFHT & MegaCam & SpARCS & u & 11.81 & 25.4 mag\\ 
\hline \hline
\end{tabular}
}
\end{table*}

An extensive range of multi-wavelength observations from ultraviolet (UV) to far-infrared (FIR) are available in the ELAIS-N1 field. An overview of the observations used in this work are given in Table \ref{tab:observations}. Here, we provide a summary of these observations and the generation of the multi-wavelength catalogue. We refer to \cite{Kondapally2020} for a detailed description of the catalogues used.

All LOFAR-detected radio sources were cross-matched with the multi-wavelength catalogues, and photometric measurements extracted, by \cite{Kondapally2020};   full details of the procedure are available in that paper. In summary, a multi-wavelength catalogue was created using forced, matched aperture photometry on pixel-matched images from all surveys. To achieve this, all individual images were resampled to a pixel scale of 0.2" and sky background is subtracted before the individual images were added together using SWarp \citep{Bertin2002ASPC..281..228B}. The flux was adjusted to a common scale using the zero-point magnitude, exposure time, and Vega-AB conversion factors, where needed. Optimal signal to noise (or $\chi^2$) detection images were created to be able to detect the faintest sources. They were created by stacking multiple bands using SWarp for the optical to near-infrared (NIR) and Spitzer Infrared Array Camera (IRAC) observations separately due to the lower resolution of the Spitzer data. The weight assigned to each band in the $\chi^2$ detection image varies according to the colour of the source. The sources were extracted from the $\chi^2$ detection image using \textsc{SExtractor} \citep{Bertin1996A&AS..117..393B}, and the two source lists were combined to make a single catalogue. The fluxes in the different bands were obtained from all sources detected in either of the $\chi^2$ detection images using aperture sizes with diameters of 1"-7" (in steps of 1") and 10". The varying point spread functions (PSF) in each filter were corrected for by aperture corrections, determined using the curve of growth estimated from moderately bright sources (see \citealt{Kondapally2020}). In this work, we used the flux measurements from the 3" apertures for the optical-NIR filters and 4" apertures for the Spitzer-IRAC filters, which have both been aperture corrected. The 3" aperture for the optical-NIR filters is less affected by PSF variations than a 2" aperture and will therefore result in more robust colours \citep{Kondapally2020}. The final ELAIS-N1 multi-wavelength catalogue contains over 2.1 million sources; 1.5 million of these sources in the overlapping PanSTARRS, UKIDSS-DXS, and Spitzer-SWIRE surveys are used for radio-optical cross-matching in an area of 6.7 deg$^2$. The cross-matching was carried out using an adaptation of the technique developed in LoTSS DR1 and presented by \cite{williams2019A&A...622A...2W}. In short, optical and IR counterparts of radio sources were identified by either the statistical likelihood ratio method or by visual classification schemes, as determined by a decision tree described in \cite{williams2019A&A...622A...2W} and \cite{Kondapally2020}. Sources with extended and/or complex radio emission were associated and classified using a combination of the LOFAR Galaxy Zoo \citep{williams2019A&A...622A...2W} and an expert-user work-flow designed for de-blending radio sources. A detailed description of the procedure followed to create the catalogues is available in \cite{Kondapally2020}. 

In addition, the Spectral and Photometric Imaging Receiver (SPIRE) and Photodetector Array Camera and Spectrometer (PACS) FIR fluxes were added to the catalogue obtained from the Herschel Extragalactic Legacy Project (HELP; Oliver et al. 2020, in prep). The procedure for the determination of the FIR fluxes is described in \cite{McCheyne2020}. In short, mid-infrared (mid-IR) and FIR fluxes were derived from the prior driven de-blending of the Spitzer Multiband Imaging Photometer (MIPS) and Herschel PACS/SPIRE imaging using the \textsc{xid+} deblending code \citep{hurley2017MNRAS.464..885H}.

Photometric redshifts for all optical sources in the \cite{Kondapally2020} catalogues were estimated using a hybrid template and machine-learning method. Full details of the method and characterisation of the redshift performance are presented in \citet[][hereafter D20]{Duncan2020}. 

\section{Characterising the radio-detected SILVERRUSH population}
\label{sec:radio_silverrush}

\begin{figure*}
\centering
   \includegraphics[width=\textwidth, trim={0.0cm 1.0cm 1.5cm 0.5cm},clip]{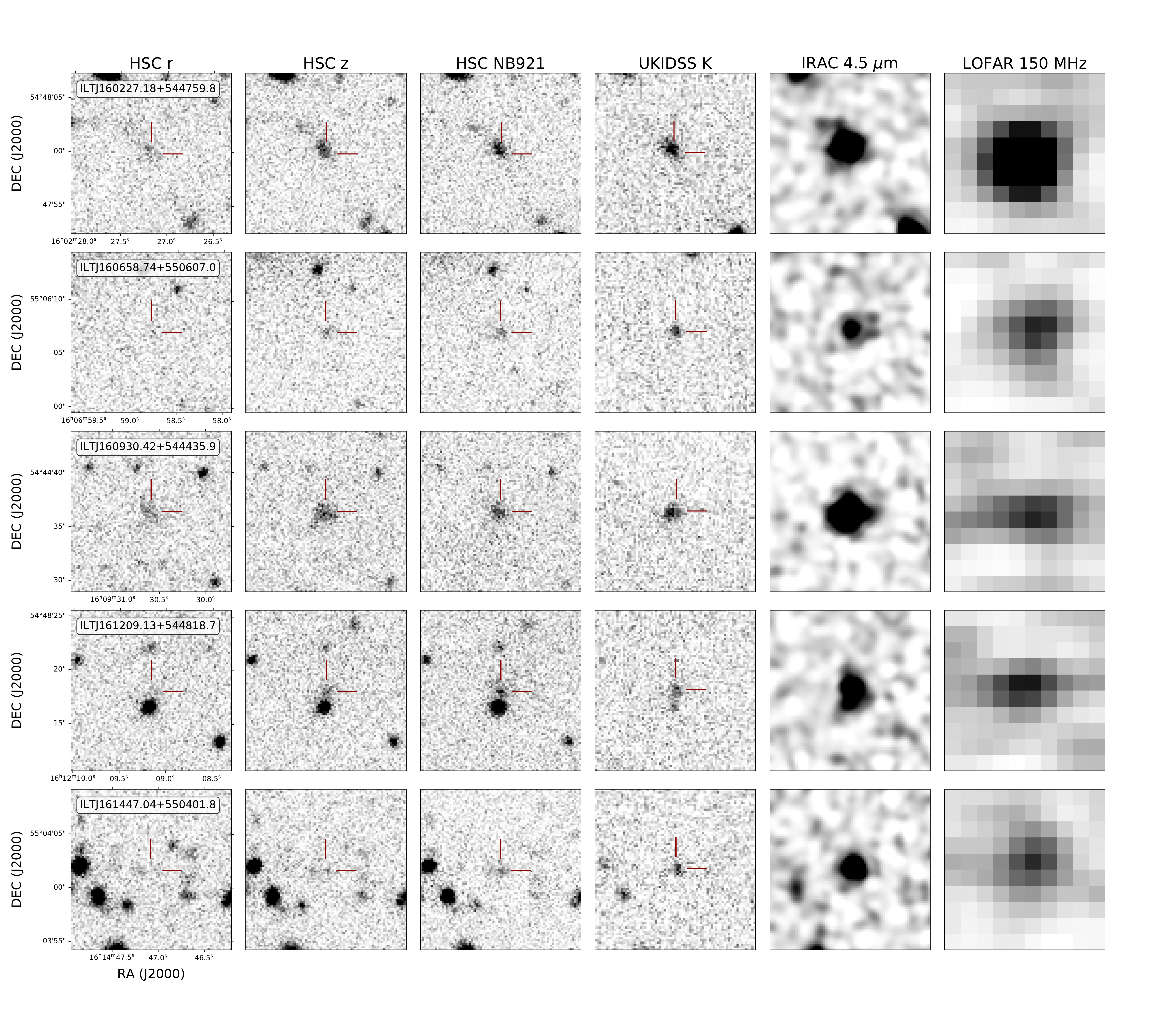}
     \caption{Multi-wavelength cutouts of the LOFAR-detected sample including, from left to right: HSC \textit{r} band, HSC \textit{z} band, HSC NB921 band, UKIDSS K band, IRAC 4.5 $\mu$m, and LOFAR 150 MHz. The sources show negligible or no emission below the Lyman break in the \textit{r} band; however, clear emission is seen in both the UKIDSS k band and IRAC. } 
     \label{fig:multiwavelength_cutouts}
\end{figure*}

The LOFAR-detected SILVERRUSH population in ELAIS-N1 was determined by cross-matching sources from the LoTSS catalogue and the SILVERRUSH NB921 and NB816 LAE catalogues within 3" separation (half of the LOFAR 6" beam size). The location of the optical counterparts of the radio sources were used for cross-matching. The resulting cross-matched sources are inspected by eye, leading to five LOFAR-detected SILVERRUSH sources out of the eight potential matches, which will from now on be referred to as the `LOFAR-detected sample'. All five of these sources matched with SILVERRUSH sources in the NB921 LAE catalogue ($z=6.6$), whereas no sources have been found to match with the NB816 LAE catalogue. The multi-wavelength cutouts of the LOFAR-detected sample are shown in Fig. \ref{fig:multiwavelength_cutouts}.

Table \ref{tab:source_properties} summarises the photometric source properties, including the photometric redshift (photo-$z$s) derived by \citetalias{Duncan2020}. Only one source (ILTJ160658.74+550607.0) has a photo-$z$ $z > 6$, whereas estimates for other sources place them at $1.4 < z < 2.1$, suggesting that they are likely to be low-redshift interlopers. It is furthermore noticeable that z $-$ NB921 $\geq$ 1.0 within 1$\sigma$ in the LoTSS photometry, but it is not as strong as the HSC values. This is likely due to the difference in underlying photometry used, as well as the large uncertainties in \textit{z} magnitude, which could partly be due to the chosen 3" aperture size. To further investigate the nature of these five sources and their probable redshifts, we examined their spectral energy distributions (SEDs). In SED fitting, we also included the FIR data, which is not included in the photo-$z$ estimates.

\begin{table*}
\caption{Multi-wavelength photometric properties and photo-$z$s of the LOFAR-selected sample. The 4.5 $\mu$m magnitude is taken from the SWIRE survey \citep{Lonsdale2003PASP..115..897L}, and the \textit{i} and \textit{z} magnitudes from HSC-SSP \citep{Aihara2018PASJ...70S...8A} are both taken from the LoTSS Deep Fields catalogue \citep{Kondapally2020}. The photometric redshifts $z_{1, \text{median}}$ and $z_{2, \text{median}}$ were obtained from LoTSS Deep Fields catalogue (see \citetalias{Duncan2020}), where $z_{1, \text{median}}$ and $z_{2, \text{median}}$ are the medians of the primary and secondary redshift peaks, respectively. A more detailed SED fitting shows that these sources are all likely [O\textsc{ii}] emitters at z=1.47 (see Sect. \ref{subsection:sed_fitting_5}).}      
\label{tab:source_properties}
\centering  
\resizebox{\textwidth}{!}{
\begin{tabular}{c c c c c c c c c }  
\hline\hline       
Source name & S$_{150\text{MHz}}$ (mJy) & 4.5 $\mu$m mag & K mag & NB921 mag & \textit{i} mag & \textit{z} mag & \multicolumn{2}{p{2cm}}{\centering {photo-$z$} } \\ 
& & & & & & & $z_1$ & $z_2$\\[0.1cm]
\hline
ILTJ160227.18+544759.8 & 0.35$\pm$0.04 & 19.95 $\pm$ 0.07 & 21.10 $\pm$ 0.11 & 23.18 $\pm$ 0.15 & 24.64 $\pm$ 0.43 & 23.76 $\pm$ 0.32  & 1.7$^{+0.8}_{-0.6}$ & 2.7$^{+0.1}_{-0.1}$ \\ [0.1cm]
ILTJ160658.74+550607.0 & 0.10$\pm$0.04 & 20.99 $\pm$ 0.17 & 22.39 $\pm$ 0.34 & 24.48 $\pm$ 0.50 & 25.60 $\pm$ 1.04 & 25.33 $\pm$ 1.37 & 6.1$^{+0.5}_{-0.8}$ & 4.4$^{+0.6}_{-0.6}$ \\[0.1cm] 
ILTJ160930.42+544435.9 & 0.11$\pm$0.05 & 20.13 $\pm$ 0.08 & 21.17 $\pm$ 0.12 & 23.19 $\pm$ 0.15 & 24.59 $\pm$ 0.41 & 23.90 $\pm$ 0.37 & 1.6$^{+0.5}_{-0.4}$ & - \\ [0.1cm]
ILTJ161209.13+544818.7 & 0.18$\pm$0.05 & 20.57 $\pm$ 0.12 & 21.65 $\pm$ 0.18 &22.26 $\pm$ 0.06 & 23.49 $\pm$ 0.15 & 23.16 $\pm$ 0.19 & 1.4$^{+0.1}_{-0.2}$ & 0.85$^{+0.06}_{-0.03}$ \\[0.1cm] 
ILTJ161447.04+550401.8 & 0.09$\pm$0.03 & 20.73 $\pm$ 0.13 & 22.16 $\pm$ 0.29 & 24.49 $\pm$ 0.50 & 25.76 $\pm$ 1.19 & 25.71 $\pm$ 1.94 & 2.1$^{+1.4}_{-1.0}$ & 4.4$^{+0.6}_{-0.6}$\\ [0.1cm]
\hline \hline
\end{tabular}}
\end{table*}

\subsection{SED fitting}
\label{subsection:sed_fitting_5}

\begin{figure*}
\centering
   \includegraphics[width=0.9\textwidth, trim={0.2cm 20cm 1cm 0.15cm}, clip]{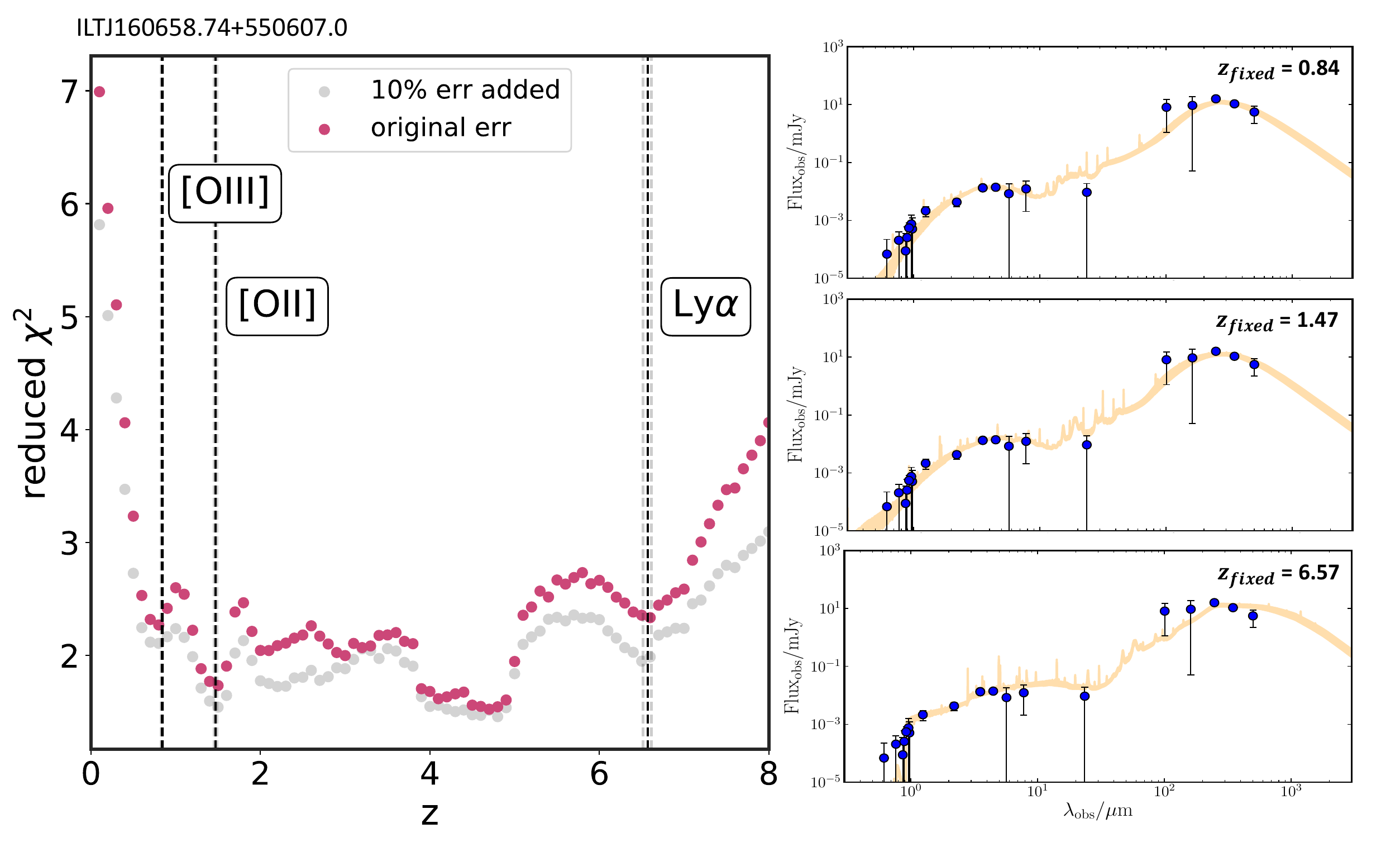}
     \caption{SED fitting results for ILTJ160658.74+550607.0, obtained using \textsc{Bagpipes}. Left: Reduced $\chi ^2$ for SED fits in the range 0 < $z$ < 8, with locations of the Ly$\alpha$, [O\textsc{ii}], and [O\textsc{iii}]+H$\beta$ emission lines indicated by dashed black lines. The dashed grey lines indicate the redshift range probed by the NB filter for the strong emission lines [O\textsc{iii}], [O\textsc{ii}], as well as Ly$\alpha$. Right: Observed flux of ILTJ160658.74+550607.0 (blue) in mJy with the original errors and posterior median model (yellow) for fits with redshifts fixed at $z=$0.84, 1.47, and 6.57. The source is best fitted by a template with z=1.47 when considering the possible emission lines. No other emission line is strong enough for the source to be at $z\sim4$.}
     \label{fig:sed_2_sources}
\end{figure*}

We performed SED fitting on the LOFAR-detected sample using the code Bayesian Analysis of Galaxies for Physical Inference and Parameter EStimation\footnote{\url{https://bagpipes.readthedocs.io/en/latest/}} (or \textsc{Bagpipes}; \citealt{Carnall2018}). In summary, \textsc{Bagpipes} is a Bayesian spectral fitting code developed for generating complex model galaxy spectra from photometric and spectroscopic observations using the MultiNest nested sampling algorithm \citep{Feroz2019OJAp....2E..10F}. The input observations can range from far-UV to microwave wavelengths. The model parameters used for fitting are the star-formation history (SFH), the nebular component, and the dust attenuation and emission components. An overview of the parameter values used in this work is given in Table \ref{tab:bagpipes_parameters}, and the available input photometric data  are summarised in Table \ref{tab:observations}. The SFH is modelled using a double power law with parameters for the total stellar mass formed and the metallicity, and the dust attenuation model of \cite{calzetti2000ApJ...533..682C} is used. A more extensive description of \textsc{Bagpipes} and its performance is available in \cite{Carnall2018}.

Besides Ly$\alpha$, other strong nebular emission lines that could account for a NB excess in the NB921 filter are H$\alpha$, [O\textsc{iii}], H$\beta$, and [O\textsc{ii}] nebular emission lines at $z\sim0.40, 0.84, 0.89$, and 1.47, respectively. The photo-$z$ accuracy is not high enough to be able to distinguish the H$\beta$ emission line from the [O\textsc{iii}] emission line, so throughout this work we refer to it as the [O\textsc{iii}] emission line at $z\sim$0.84 since this line is more prevalent in emission line galaxies (e.g. \citealt{Hayashi2018PASJ...70S..17H, sobral2015MNRAS.451.2303S, Khostovan2015MNRAS.452.3948K}). The combination of the implausibly high dust extinction required to produce such red SEDs at $z=0.4$ and the proportionally low co-moving volume probed means that the contamination of this sample by H$\alpha$ emitters at $z = 0.4$ is thought to be negligible. Therefore, specific models for [O\textsc{iii}], [O\textsc{ii}], and Ly$\alpha$ emission lines are fitted to the LAE candidates. 

It must be noted that \textsc{Bagpipes} does not include AGN templates, which could lead to an underestimation of the mid-IR flux. An AGN contribution can be therefore be identified by the inability of \textsc{Bagpipes} to fit the mid-IR flux. The possible effects of the lack of AGN templates are therefore discussed together with the results.

To investigate the probability of a source residing at the above-mentioned redshifts, the SED fits are performed with fixed $z=0.84,1.47,6.57$ as well as at 80 redshifts in the range $z=[0.1,8]$, in steps of 0.1. For each redshift, the SED fit resulting in the minimum reduced chi-squared value ($\chi^2_{\nu}$) is determined, to be used as a comparison for the different models. Additionally, non-detections, with S/N $<$ 3.0, are used by \textsc{Bagpipes} and in the $\chi^2_{\nu}$ calculation. However, negative fluxes are not taken into account. The SED fitting results of the high-z LAE candidate ILTJ160658.74+550607.0 are shown in Fig. \ref{fig:sed_2_sources}. The $\chi^2_{\nu}$ value lies within a small range and is lowest at $z\sim4.9$ when fitting with the original errors. However, in panchromatic SED fittings, there are many model uncertainties and the flux errors arising from flux calibration issues or correlated noise are usually underestimated (see e.g. \citealt{Marquez2014A&A...563A..14M}). We therefore re-fitted the LOFAR-detected sample, incorporating an additional 10 \% flux uncertainty in quadrature to assess what effect this may have on the preferred redshift solution (see Fig. \ref{fig:sed_2_sources}). The overall shape of the curve remains largely the same with the additional 10\% flux uncertainty, including the decrease around $z\sim4-5$, though it does improve the fit at $z=1.47$, making it comparable within the errors to the $z\sim4-5$ fit. However, around this redshift of $z\sim4-5,$ there is no emission line with a typical equivalent width (EW) high enough to be detected in the HSC NB921 filter. Also, because of the large number of data points, the relative contribution of the NB measurement to the overall $\chi^2$ value is limited. The decrease in $\chi^2$ can therefore be attributed to the BB SED shape. In the right-hand panel of Fig. \ref{fig:sed_2_sources}, the best-fitting posterior median models at $z=0.84, 1.47,$ and 6.57 for [O\textsc{iii}], [O\textsc{ii}], and Ly$\alpha$, respectively, are shown in yellow. Here, the SED fit at $z=1.47$ is the preferred solution. The SED fitting results of the four other sources in the LOFAR-detected sample are shown in Fig. \ref{fig:sed_3_sources}. The $\chi^2$ values are minimised around $z\sim1.5$, providing strong evidence that these are [O\textsc{ii}] emitters at $z=1.47$.

To compare the preference for each model, we used the Bayesian information criterion (BIC), which in its general form is given by 
\begin{equation}
    \text{BIC} = -2\ln{L}+ k\ln{N},
\end{equation}
where L is the maximised likelihood, k the number of parameters, and N the number of observations (see \citealt{kass1995bayes}). In the case of Gaussian distributed model errors, this becomes
\begin{equation}
   \text{BIC} = \chi^2 + k\ln{N}.
\end{equation}
The model with the highest probability minimises this value, and, when comparing models, a $\Delta$BIC > 2 gives positive evidence for one of the models being preferred over the other, whereas -2 < $\Delta$BIC < 2 indicates that there is no evidence for either model being preferred. A $\Delta$BIC > 6 gives strong positive evidence (as defined by \citealt{kass1995bayes}). When comparing our SED fits, the number of observations and parameters remains constant; therefore, BIC becomes $\chi^2$ and we can use the $\Delta \chi^2$ values of the fits at $z=0.84$, $1.47$, and $6.57$ to determine the preference for each model. The $\chi^2_{\nu}$ and $\Delta \chi^2$ values of the Ly$\alpha$, [O\textsc{ii}], and [O\textsc{iii}] emission line fits to the LOFAR-selected sample are summarised in Table \ref{tab:delta_chisquared}. Here, the original errors are used to compare the models since the 10\% added flux error causes over-fitting of the data, resulting in $\chi^2_{\nu}$ values below 1. For all five sources, $\Delta \chi^2 (\text{Ly}\alpha-\text{[O\textsc{ii}]})$ and/or $\Delta \chi^2 (\text{Ly}\alpha-\text{[O\textsc{iii}]})$ > 2, meaning the models for [O\textsc{ii}] or [O\textsc{iii}] provide a better fit to the data and Ly$\alpha$ is likely not to be the detected emission line. In addition, $\Delta \chi^2 (\text{[O\textsc{iii}]}-\text{[O\textsc{ii}]})$ > 2 for all sources, suggesting the detected emission line is [O\textsc{ii}], in turn implying that the sources are situated at $z=1.47$. This conclusion is also supported by comparing the co-moving volume being probed at z=0.84 and z=1.47, which is $\sim$2 times higher for $z=1.47$ when assuming a flat $\Lambda$-$CDM$ cosmology. The evidence for Ly$\alpha$ not being the source of emission is less strong in the case of ILTJ160658.74+550607.0 (shown in Fig. \ref{fig:sed_2_sources}); however, the FIR SED shape, consisting of SPIRE and PACS observations, is best reproduced by the $z=1.47$ model and therefore seems to be the actual redshift. Follow-up observations are needed to confirm the redshift of this source. If this source is a radio galaxy at $z=6.6$, it would be the most distant radio galaxy known to date.

\begin{table*}
\caption{Reduced $\chi^2$ and $\Delta \chi^2$ values from SED fitting (using original error) to redshift $z=6.57\pm0.05$, $1.47\pm0.02$, and $0.84\pm0.02$, corresponding to emission lines Ly$\alpha$, [O\textsc{ii}], and [O\textsc{iii}], respectively, taking into account the FWHM (133.45 \r{A}) of the NB921 HSC filter. The $\Delta \chi^2$ values suggest that the five sources are [O\textsc{ii}] emitters at $z=1.47$. } 
\label{tab:delta_chisquared}      
\centering
\begin{tabular}{c c c c c c c}  
\hline\hline       
Source name & \multicolumn{3}{p{5cm}}{\centering {Reduced $\chi^2$} } & \multicolumn{3}{p{5cm}}{\centering {$\Delta$ $\chi^2$} } \\

& Ly$\alpha$ & [O\textsc{ii}] &  [O\textsc{iii}] &  Ly$\alpha$ $-$ [O\textsc{ii}] &  Ly$\alpha$ $-$ [O\textsc{iii}] & [O\textsc{iii}] $-$ [O\textsc{ii}]\\ [0.1cm]
\hline
ILTJ160227.18+544759.8 & 5.0$^{+3.1}_{-0.6}$ & 0.9$^{+0.0}_{-0.1}$ & 2.0$^{+0.2}_{-0.0}$ & 53-95 & 33-77 & 18-21\\ [0.1cm]
ILTJ160658.74+550607.0 & 2.3$^{+0.2}_{-0.0}$ & 1.7$^{+0.1}_{-0.1}$ & 2.3$^{+0.1}_{-0.0}$& 5.7-8.6 & 0.1-2.0 & 5.8-7.5\\ [0.1cm]
ILTJ160930.42+544435.9 & 4.6$^{+1.69}_{-0.2}$ & 0.9$^{+0.1}_{-0.1}$ & 2.1$^{+0.1}_{-0.1}$&  60-93& 38-73 &19-23\\[0.1cm] 
ILTJ161209.13+544818.7 & 25.6$^{+3.3}_{-2.7}$ &  0.6$^{+0.5}_{-0.0}$ & 5.5$^{+0.2}_{-0.3}$ & 371-476 & 293-401 &71-83\\ [0.1cm]
ILTJ161447.04+550401.8 & 1.1$^{+0.2}_{-0.1}$& 0.4$^{+0.0}_{-0.0}$ &  0.79$^{+0.0}_{-0.05}$ & 8.4-11.4 & 3.5-7.1 &4.3-4.9\\[0.1cm] 
\hline \hline
\end{tabular}
\end{table*}

We present the derived stellar mass and star-formation rate for each fixed-redshift SED fit in Table \ref{tab:sed_parameters}. When redshift is fixed at $z=6.57$, the derived stellar masses are 10$^{11.7}$ < M$_*$ < 10$^{11.9}$ M$_{\odot}$ and SFRs  are$\sim$650$-$4400 M$_{\odot}$ yr$^{-1}$. Even given the large uncertainties on the high mass end of the stellar mass function at these redshifts, the likelihood of masses being this high is small (see e.g. \citealt{Duncan2014, Grazian2015A&A...575A..96G, Song2016ApJ...825....5S}). The median Ly$\alpha$ SFR obtained by, for example, \cite{calhau2020MNRAS.tmp..448C} is 9.8$^{+9.7}_{-5.2}$ M$_{\odot}$ yr$^{-1}$ for their sample at $z > 3.5$, when excluding AGN. The SFRs derived from their FIR-detected LAEs (of which 76\% are X-ray or radio AGN) using their IR fluxes yield higher values of 200$^{+430}_{-110}$ M$_{\odot}$ yr$^{-1}$, with a few sources > 600 M$_{\odot}$ yr$^{-1}$.

As noted before, \textsc{Bagpipes} does not include AGN templates. If there were a strong AGN contribution, there should be a warm dust contribution at mid-IR wavelengths. However, none of these five sources show excess in the MIPS 24$\mu$m band. In addition, all these sources are classified as star-forming from both their optical and radio emissions via the hybrid SED fitting method from \cite{Best2020}. Furthermore, in the photo-$z$ fitting procedure of \citetalias{Duncan2020}, AGN templates are used, and these results are consistent with the redshifts obtained from SED fitting. Unlike in \citetalias{Duncan2020}, in the \textsc{Bagpipes} modelling FIR fluxes have been included in the fit, which, for the source ILTJ160658.74+550607.0, helps break the degeneracy between the $z=1.5$ and $6.6$ solutions. The remaining small differences in the resulting redshifts between this work and the photo-$z$s of \citetalias{Duncan2020} can be attributed to the use of a different set of templates (see \citealt{duncan2018MNRAS.473.2655D}) and the machine-learning contribution in \citetalias{Duncan2020}. 

\begin{table*}
\caption{Median (50th percentile) stellar mass and SFR obtained from the SED fitting of the LOFAR-selected sample, with fixed redshifts at $z=0.84, 1.47,$ and 6.57. The errors are given by the 84th and 16th percentiles.}      
\label{tab:sed_parameters}      
\centering  
\resizebox{\textwidth}{!}{
\begin{tabular}{c  | c c | c c | c c }  
\hline\hline
 &  \multicolumn{2}{c}{$z=0.84$} & \multicolumn{2}{c}{$z=1.47$} & \multicolumn{2}{c}{$z=6.57$} \\
\hline
Source name & log $M_*$ (M$_\odot$) & SFR (M$_\odot$ yr$^{-1}$) & log $M_*$ (M$_\odot$) & SFR (M$_\odot$ yr$^{-1}$) & log $M_*$ (M$_\odot$) & SFR (M$_\odot$ yr$^{-1}$) \\[0.2cm]
\hline
ILTJ160227.18+544759.8 & 10.61$^{+0.09}_{-0.11}$ & 18$^{+5}_{-2}$  & 11.12$^{+0.08}_{-0.27}$ & 82$^{+26}_{-15}$ & 11.78$^{+0.18}_{-0.14}$ & 4.4$^{+0.8}_{-0.6} \times 10^3$\\ [0.1cm]
ILTJ160658.74+550607.0 & 10.38$^{+0.12}_{-0.18}$ & 10$^{+2}_{-1}$ & 10.75$^{+0.18}_{-0.17}$ & 54$^{+13}_{-10}$ & 11.69$^{+0.20}_{-0.23}$ & 2.2$^{+0.3}_{-0.3} \times 10^3$ \\[0.1cm]
ILTJ160930.42+544435.9 & 10.65$^{+0.04}_{-0.09}$ & 15$^{+1}_{-2}$ & 11.13$^{+0.03}_{-0.06}$ & 61$^{+8}_{-8}$ & 11.87$^{+0.24}_{-0.12}$ & 2.6$^{+0.4}_{-0.3} \times 10^3$\\[0.1cm]
ILTJ161209.13+544818.7 & 10.25$^{+0.04}_{-0.06}$ & 6$^{+1}_{-1}$ & 10.70$^{+0.04}_{-0.08}$ & 24$^{+4}_{-3}$ & 11.68$^{+0.06}_{-0.07}$ & 8.8$^{+0.2}_{-0.2}\times 10^2$\\[0.1cm]
ILTJ161447.04+550401.8 & 10.38$^{+0.14}_{-0.18}$ & 10$^{+6}_{-2}$ & 10.79$^{+0.10}_{-0.12}$ & 31$^{+11}_{-7}$ & 11.91$^{+0.13}_{-0.13}$ & 6.5$^{+0.7}_{-0.6}\times 10^2$ \\[0.1cm]
\hline \hline
\end{tabular}}
\end{table*}

\subsection{Emission line and radio properties}
\label{subsec:emission_line_radio}

From the photometric data, we determined the physical properties of the LOFAR-selected sample, such as radio luminosity, line luminosity, and EW (see Table \ref{tab:physical_properties}). As each of these quantities depend on the redshift of the source, the values are given for fixed $z=0.84, 1.47,$ and 6.57. A spectral index $\alpha$ of -0.7$\pm$0.7 (defined by $S_{\nu} \propto \nu^{\alpha}$ with $S_{\nu}$ the flux density in Jy) was assumed when determining the radio luminosity, where the error is based on the spectral index distribution derived by \cite{Saxena2018}. No spectral index could be obtained for the sources in the LOFAR-selected sample
since no sufficiently deep radio data (e.g. from the Very Large Array  or the Giant Metrewave Radio Telescope) was publicly available. For the calculation of the EW and line luminosity, we used the equations presented in \cite{sobral2012}. Subsequently, the line luminosity and radio luminosity were calculated using the line flux and radio flux, respectively. The rest-frame EW (EW$_0$) was determined by dividing the observed EW by $1+z$ to correct for the Hubble expansion. The values obtained for each assumed redshift are shown in Table \ref{tab:physical_properties}.

A large range of Ly$\alpha$ rest EW values are quoted in the literature. Ly$\alpha$ EW values of, for example,  $\sim$0$-$400 $\r{A}$ \citep{Ono2010MNRAS.402.1580O} and 10$-$1000 $\r{A}$ \citep{leclercq2017A&A...608A...8L} have been found. Furthermore, the Ly$\alpha$ LFs presented in \cite{calhau2020MNRAS.tmp..448C} and \cite{Konno2018_SR4} show Ly$\alpha$ luminosities ranging from $\sim$10$^{42.5} - $ 10$^{44.0}$ erg s$^{-1}$. A global study of LAEs at $2.5 < z < 5.8$ by \cite{Sobral2018MNRAS.476.4725S} in the COSMOS field resulted in a characteristic luminosity L$_{\text{Ly}\alpha}$ $\sim$ 10$^{43.8 \pm 0.1}$ erg s$^{-1}$ of the Schechter function. The values for the EW and L$_{\text{Ly}\alpha}$ found in this work are consistent with these literature results.

The EW$_0$ derived when we assumed that the detected lines were [O\textsc{ii}] and [O\textsc{iii}] are comparable to the EW distribution found in \cite{Khostovan2016MNRAS.463.2363K}, where the majority of the [O\textsc{iii}]+H$\beta$ line emitters at $z=0.84$ have 20 < EW$_0$ < 500, and the majority of [O\textsc{ii}] line emitters at $z=1.47$ have 20 < EW$_0$ < 300. The derived [O\textsc{ii}] and [O\textsc{iii}] luminosities are within the ranges of 41.4 < log$_{10}$(L$_{\text{[O\textsc{iii}]}}$) < 42.3 and 40.5 < log$_{10}$(L$_{\text{[O\textsc{iii}]}}$) < 42.1. However, it must be noted that the sources in this study could be different from the general population of [O\textsc{ii}] and [O\textsc{iii}] emitters since this sub-sample of sources satisfies the LAE selection criteria. These criteria (see Eqs. \ref{eq:idrop} and \ref{eq:zdrop}) select for sources with high magnitude differences between the \textit{z} band and NBs, as well as non-detections in filters below the supposed Lyman break, suggesting that these are potentially sources with strong Balmer breaks.

The obtained radio luminosities and SFRs from \textsc{Bagpipes} can be compared to the L$_{150\rm{MHz}}$-SFR relation derived by \cite{gurkan2018MNRAS.475.3010G}, assuming the relation does not evolve with redshift. This assumption is supported by the study of \cite{Duncan2020b}, where no redshift evolution is found out to z$\sim$2.6. If the sources are located at $z=1.5$, all five sources are slightly offset to higher radio luminosities than would be expected from the L$_{150\rm{MHz}}$-SFR relation, with a difference of 0.04-0.69 dex from the 1$\sigma$ upper limit, which indicates that they could be low luminosity AGN. However, these deviations are non-significant when taking the 0.3 dex error on the radio luminosities and the $\sim$0.3 dex scatter of observations in the L$_{150\rm{MHz}}$-SFR relation into account. Moreover, the SED fitting did not suggest an AGN contribution. The \cite{gurkan2018MNRAS.475.3010G} relation is consistent with the new derived L$_{150\rm{MHz}}$-SFR relation from the LOFAR Deep Fields data by \cite{Smith2020} .

\begin{table*}
\caption{Derived total radio luminosity L$_{150\rm{MHz}}$, line luminosity L, and rest-frame EW$_0$ of the LOFAR-selected sample for SED fits with redshifts fixed at $z=0.84, 1.47,$ and 6.57, using the LOFAR 150 MHz total flux measurements and HSC flux values in the LoTSS catalogue.}      
\label{tab:physical_properties}      
\centering  
\resizebox{\textwidth}{!}{
\begin{tabular}{c | c c c | c c c | c c c }  
\hline\hline
 & \multicolumn{3}{c}{$z=0.84$} & \multicolumn{3}{c}{$z=1.47$} & \multicolumn{3}{c}{$z=6.57$} \\
\hline
Source name & \multicolumn{1}{p{2.0cm}}{\centering {log$_{10}$ $L_{150\rm{MHz}}$} \\ {(W Hz$^{-1}$)} } & \multicolumn{1}{p{2cm}}{\centering {log$_{10}$(L$_{\text{[O\textsc{iii}]}}$)} \\ {(erg s$^{-1}$)}} &   \multicolumn{1}{p{1.0cm}}{\centering {EW$_{0,\text{[O\textsc{iii}]}}$} \\ {($\r{A}$)}}
& \multicolumn{1}{p{2cm}}{\centering {log$_{10}$ $L_{150\rm{MHz}}$} \\ {(W Hz$^{-1}$)} } & \multicolumn{1}{p{2cm}}{\centering {log$_{10}$(L$_{\text{[O\textsc{ii}]}}$)} \\ {(erg s$^{-1}$)}} & \multicolumn{1}{p{1.0cm}}{\centering {EW$_{0,\text{[O\textsc{ii}]}}$} \\ {($\r{A}$)}}
& \multicolumn{1}{p{2cm}}{\centering {log$_{10}$ $L_{150\rm{MHz}}$} \\ {(W Hz$^{-1}$)} } & \multicolumn{1}{p{2cm}}{\centering {log$_{10}$(L$_{\text{Ly}\alpha}$)}\\ {(erg s$^{-1}$)}} & \multicolumn{1}{p{1.0cm}}{\centering {EW$_{0,\text{Ly}\alpha}$} \\ {($\r{A}$)}}
\\
\hline
ILTJ160227.18+544759.8 & 24.0$\pm$0.2 & 41.1$\pm$0.2 & 71$\pm$21 & 24.6$\pm$0.3 & 41.7$\pm$0.2 &  53$\pm$16& 26.0$\pm$0.6 & 43.3$\pm$0.2 & 17$\pm$5 \\
ILTJ160658.74+550607.0 & 23.5$\pm$0.3 & 40.7$\pm$0.6 & 132$\pm$151 & 24.0$\pm$0.3 & 41.3$\pm$0.6 & 99$\pm$113 & 25.4$\pm$0.6 & 42.9$\pm$0.6 & 32$\pm$36 \\
ILTJ160930.42+544435.9 & 23.5$\pm$0.3 & 41.2$\pm$0.2 & 96$\pm$27 & 24.0$\pm$0.3 & 41.8$\pm$0.2 & 72$\pm$20& 25.5$\pm$0.6 & 43.4$\pm$0.2 & 23$\pm$7  \\
ILTJ161209.13+544818.7 & 23.7$\pm$0.2 & 41.6$\pm$0.1 & 152$\pm$26 & 24.3$\pm$0.3 &42.2$\pm$0.1& 114$\pm$19& 25.7$\pm$0.6 & 43.8$\pm$0.1& 37$\pm$6  \\
ILTJ161447.04+550401.8 & 23.4$\pm$0.2 & 40.8$\pm$0.8 &  304$\pm$899 & 24.0$\pm$0.3 & 41.4$\pm$0.8 & 226$\pm$670 & 25.4$\pm$0.6 &  43.0$\pm$0.8& 74$\pm$219 \\
\hline \hline
\end{tabular}}
\end{table*}

\section{Radio and optical-IR properties of the wider LAE population}
\label{sec:radio_wider_pop}

The available evidence suggests that the LAE candidates in the LOFAR-selected sample are not located at $z\sim6.6$. Therefore, we cannot give an upper limit on the AGN fraction of LAEs at $z > 6$. These results, however, highlight different possible problems of the impurity of LAE samples as well as the low reliability of measurements on the LAE properties and the claimed AGN fractions at these redshifts. To investigate the scale of this problem, we further analysed the wider $z\sim5.7$ and $z\sim6.6$ LAE samples using our full panchromatic photometry. 

\begin{figure}
\centering
   \includegraphics[width=\columnwidth, trim={0.0cm 0.0cm 1.0cm 1.0cm}, clip]{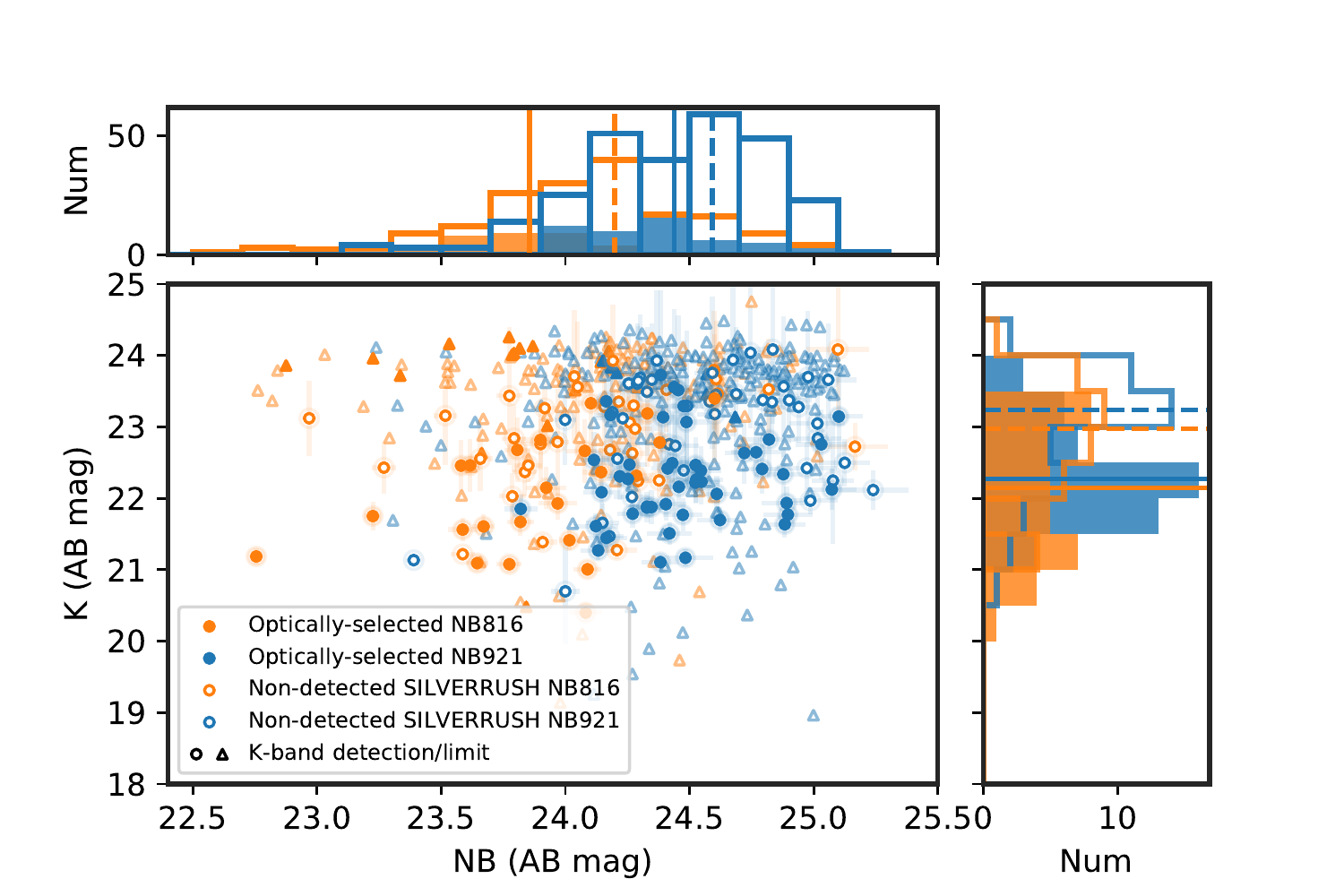}
     \caption{K and NB magnitudes (NB816 or NB921) from the optically selected sample and the non-detected SILVERRUSH sample. Limiting magnitudes (1$\sigma$) are given for sources below 1$\sigma$ detection (filled and unfilled triangles for sources in the optically selected and non-detected SILVERRUSH samples, respectively). The histograms only include sources with a >$1\sigma$ detection. The median K and NB magnitudes are indicated by the solid and dashed lines for the optically selected and non-detected samples, respectively. In the optically selected sample, 23\% of sources are not detected in K above 1$\sigma$, while for the non-detected SILVERRUSH sample, this fraction is 82\%. The optically selected sample is brighter and redder than the original SILVERRUSH sample.}
     \label{fig:magnitude_compare}
\end{figure}

The compiled optical and IR photometric catalogue for ELAIS-N1 also allows for the study of the multi-wavelength properties of SILVERRUSH sources that do not have radio detections. The sources from the LoTSS multi-wavelength catalogue and the SILVERRUSH catalogues are cross-matched within 1" and inspected by eye. Of the 349 and 229 LAEs in ELAIS-N1 in the SILVERRUSH NB921 and NB816 LAE catalogues, 58 and 53 sources, respectively, are detected in the LoTSS multi-wavelength catalogue. This sample will from now on be referred to as the `optically selected sample'. The LoTSS catalogue, selected from the $\chi^2$ detection image, may not be as optimised for Ly$\alpha$ candidates as the SILVERRUSH extraction; furthermore, SILVERRUSH used the deeper internal intermediate data release HSC-SSP 16A, so faint sources (\textit{i} mag > 26) are often missed in the LoTSS catalogue. We performed forced photometry on all SILVERRUSH sources to compare the two samples. The K and NB magnitudes from the optically selected and non-detected SILVERRUSH sample are shown in Fig. \ref{fig:magnitude_compare}. The non-detected SILVERRUSH sources with a flux measurement at $>1\sigma$ have fainter K band magnitudes (K = 23.0/23.2 for NB816/NB921) compared to the optically selected sources (K = 22.1/22.3 for NB816/NB921). The median NB magnitudes are only slightly fainter for the non-detected SILVERRUSH sources (NB = 24.2/24.6 for NB816/NB921) compared to the optically selected sources (NB = 23.9/24.4 for NB816/NB921). This indicates that the optically selected sample contains sources that are brighter and redder than the non-detected SILVERRUSH sample. We note that these magnitude differences are estimates because of the faintness of the sources and the choice of a 1$\sigma$ detection limit. The bias of our sample is further discussed in Sect. \ref{subsec:lae_selection_discussion}. 

Properties of the optically selected sample obtained from photometry and SED model fitting are discussed in Sect. \ref{subsec:sed_fit_full_pop}. Furthermore, we stack the radio and the IR and optical images of the full sample to determine the limiting magnitudes, which are presented in Sect. \ref{subsec:stacking}.

\subsection{SED fitting optically selected sample}
\label{subsec:sed_fit_full_pop}

\begin{figure*}
    \centering
    \begin{subfigure}{\columnwidth}
        \centering
        \includegraphics[width=\textwidth, trim={0cm 0.3cm 0cm 0.0cm}, clip]{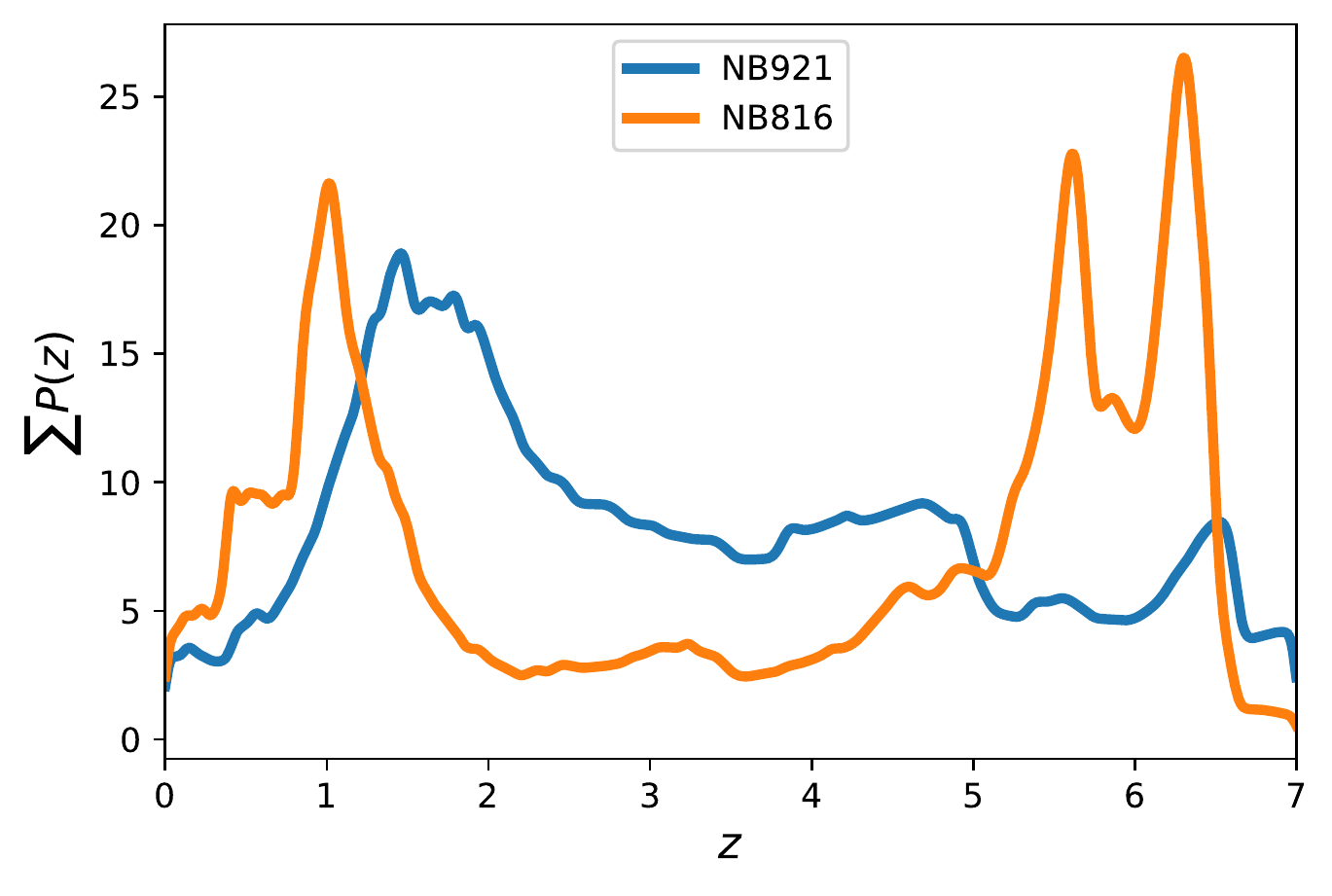}
    \end{subfigure}
     ~ 
    \begin{subfigure}{\columnwidth}
        \centering
        \includegraphics[width=\textwidth, trim={0cm 0cm 1.0cm 1.5cm}, clip]{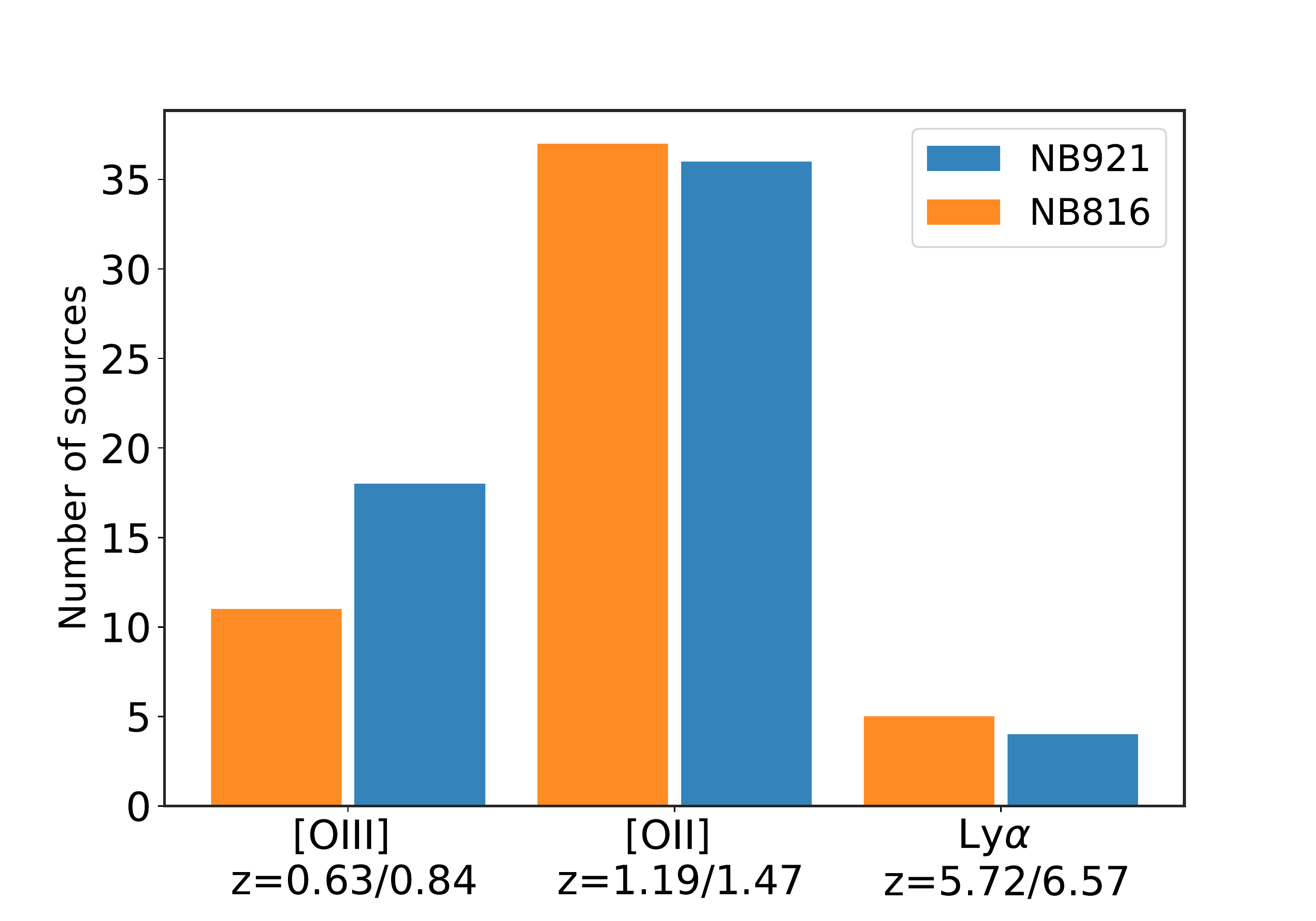}
    \end{subfigure}
    \caption{\label{fig:red_distributions} Left: Stacked photometric redshift probability distribution (sum of individual sources) of the optically cross-matched sample of the NB921 and NB816 LAE catalogues from \citetalias{Duncan2020}. Right: Histograms of the best-fitting solutions for SED fits performed with fixed-redshift solutions corresponding to the three candidate emission lines. The known possible redshifts from the NB emission line detections and more detailed SED fitting allows us to exclude other high-$z$ solutions that are sometimes favoured by the photometric redshift code.}
\end{figure*}

The stacked full photometric redshift probabilities from \citetalias{Duncan2020} for the optically selected sample are shown in the left-hand panel of Fig. \ref{fig:red_distributions}. This is a sum of the photo-zs of the individual sources. Of the two probability distributions, 44\% of the posterior is at $z>5$ for the NB816 sources, and less than 10\% of the posterior is at $z>6$ for the NB921 sources. This already suggests a high contamination fraction in both of these samples. 

\begin{figure*}
\centering
   \includegraphics[width=\textwidth, trim={0.0cm 0cm 0.0cm 0.0cm}, clip]{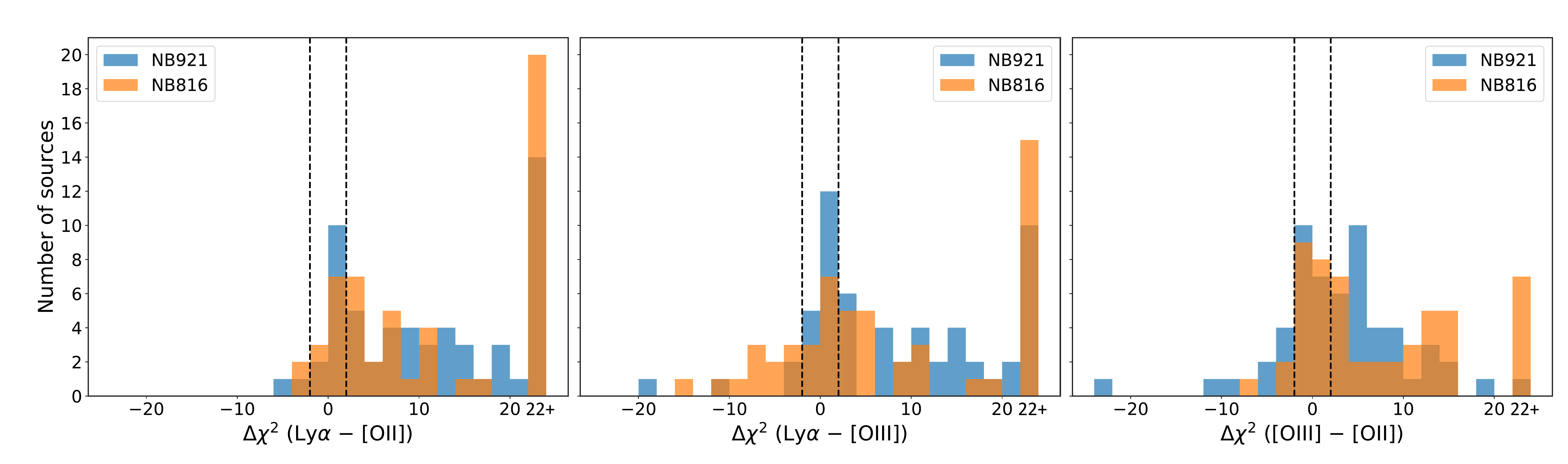}
     \caption{Comparison of $\chi^2$ values obtained from SED fitting models of Ly$\alpha$, [O\textsc{ii}], and [O\textsc{iii}] for the optical cross-matched LAE sample of NB921 and NB816. The dashed lines indicate $\Delta \chi^2$ values of 2 and $-2$. For values higher than 2 and lower than $-$2, there is a preferred model, but in between the dashed lines there is no preference for either model. The bin width is set to 2.}
     \label{fig:delta_chisqr_all} 
\end{figure*}

These photo-$z$s could potentially be biased towards lower redshifts due to the applied magnitude prior, which favours low-z solutions (z < 2) even for faint sources while not disallowing high-z solutions \citep{duncan2018MNRAS.473.2655D}. We therefore extend this with more detailed SED fitting by repeating the procedure in Sect.~\ref{subsection:sed_fitting_5} for the full optically selected sample of LAEs by fixing the redshifts at $z=0.63, 1.19, 5.72$ for the NB816 catalogue and $z=0.84, 1.47, 6.57$ for the NB921 catalogue. Of the 58 candidate LAEs at $z=5.7$, nine have FIR detections. For the $z=6.6$ candidates, six out of 53 have FIR detections. The other sources have only UV-NIR data. We fitted all available data for each source. For each source, we identified the strongest emission line with the best fitting SED (lowest $\chi^2_{\nu}$) and assigned the sources the corresponding redshift. We present the numbers of sources at each of the redshifts in Fig. \ref{fig:red_distributions} (right-hand panel). For five sources in NB921 and one source in the NB816 LAE catalogue, the SED fits are poorly constrained due to detections in fewer than nine photometric bands, hence these results are excluded from further analysis. The best SED fits (right-hand panel in Fig. \ref{fig:red_distributions}) suggest contamination rates of 90\%\ and 93\% for the NB816 and NB921 catalogue matches, respectively. To quantify this further, we repeated the model comparison method for this larger non-radio-detected sample, again with fixed redshifts. The resulting $\Delta \chi^2$ values for $\Delta \chi^2$ (Ly$\alpha$ $-$[O\textsc{ii}]), $\Delta \chi^2$ (Ly$\alpha$ $-$[O\textsc{iii}]), and $\Delta \chi^2$ ([O\textsc{iii}]$-$[O\textsc{ii}]) are presented in Fig. \ref{fig:delta_chisqr_all}, where the dotted lines indicate $\Delta \chi^2$ values of 2 and $-$2. Sources likely to be Ly$\alpha$ emitters at $z=$6.57/5.72 satisfy $\Delta \chi^2$ < $-$2 for either $\Delta \chi^2$(Ly$\alpha$ $-$[O\textsc{ii}]) or $\Delta \chi^2$(Ly$\alpha$ $-$[O\textsc{iii}]), under the condition that the other $\Delta \chi^2$ $\ngeq$ 2. This is the case for four sources in the NB816 and five sources in the NB921 catalogue. On the other hand, 43 and 47 of the sources have $\Delta \chi^2$ > 2 for either of the two $\Delta \chi^2$ values, for both NB816 and NB921, so these are more likely to be $z=$0.63/0.84 or $z=$1.19/1.47 contaminants, respectively,  rather than $z=$5.72/6.57 LAEs. The model comparison for sources with $\Delta \chi^2$ between $-$2 and 2 is inconclusive and can be regarded as an estimate of the uncertainty. The contamination rate determined from SED fittings using the three emission lines [O\textsc{ii}], [O\textsc{iii}], and Ly$\alpha$ is therefore 81-92\% (43-49 out of the 53 sources) and 81-91\% (47-53 out of the 58 sources) for NB816 and NB921, respectively. Similarly, comparing the $\chi^2$ of [O\textsc{ii}] and [O\textsc{iii}], [O\textsc{ii}] is likely ($\Delta \chi^2 > 2$) to be the emission line for 63$\pm$31\% and 55$\pm$28\% of the possible contaminating sources for NB816 and NB921, respectively, whereas [O\textsc{iii}] is likely to be the corresponding emission line for 6$\pm$31\% and 17$\pm$28\% of the sources for NB816 and NB921, respectively (with the remainder inconclusive).
Comparing these results to the primary and secondary photo-zs of \citetalias{Duncan2020}, we obtain that, within the error, 21/53 (40\%) of the photo-zs in the NB816 sample and 41 of the 58 (71\%) photo-zs in the NB921 sample are consistent. The remaining discrepancy between the two photo-z estimates can be attributed to the faint magnitudes of the sources and the high redshifts, which significantly increase the photo-z uncertainties and result in a broad photo-z posterior (see \citetalias{Duncan2020}). Comparing our contamination rates to the stacked photo-z probability distribution in the left-hand panel of Fig. \ref{fig:red_distributions} shows that the photo-z estimates are consistent for the NB921 sample, but contamination is estimated to be lower for the NB816 sample. The photometric redshifts of \citetalias{Duncan2020} and the SED fitting results together therefore suggest the contamination of our subset is 56-92\% and 81-91\% for the NB816 and NB921 samples, respectively.

\begin{figure*}
\centering
   \includegraphics[width=\textwidth, trim={0.0cm 0cm 0cm 0.0cm}, clip]{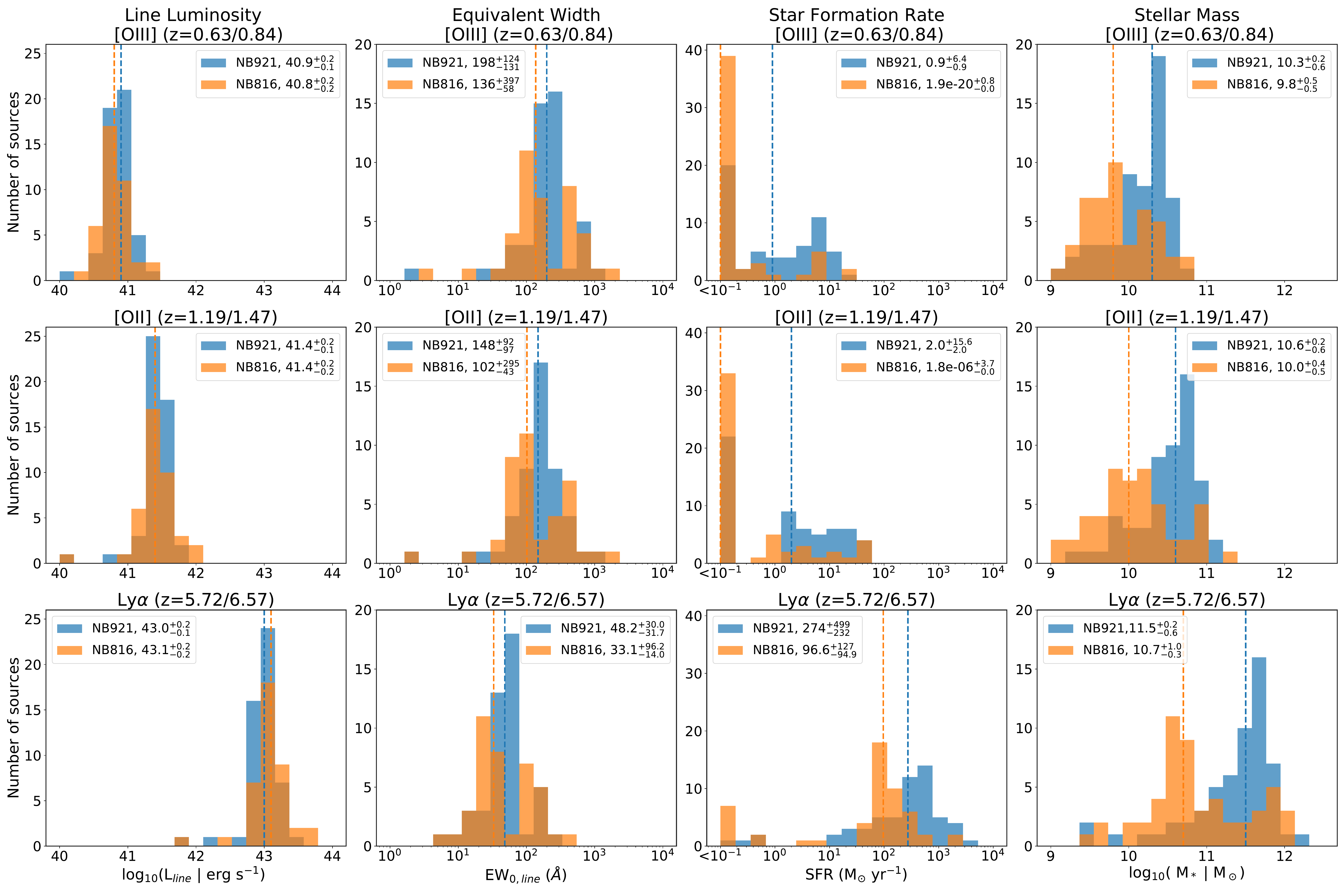}
     \caption{Derived line luminosity, EW, SFR, and stellar mass of the optical cross-matched LAE sample with the NB921 and NB816 LAE catalogues at different fixed redshifts of $z=$0.63/0.84, 1.19/1.47, and 5.72/6.57. The dashed blue and orange lines indicate the median for NB921 and NB816, respectively; the values given in the legend include errors derived from the central 68\% range of the distribution. The lowest bin in the SFR histograms show the sources with SFR < 10$^{-1}$ M$_{\odot}$ yr$^{-1}$.}
     \label{fig:lae_poperties}
\end{figure*}

Physical properties of the samples, derived from SED fitting (SFR and stellar mass) and from the photometric data (line luminosity and EW) are shown in Fig. \ref{fig:lae_poperties} for each of the three proposed redshifts. Assuming these sources are LAEs, the median Ly$\alpha$ luminosity is 10$^{43.1\pm0.2}$ and 10$^{43.0^{+0.2}_{-0.1}}$ erg s$^{-1}$ for NB816 and NB921, in line with other works from, for example, \cite{calhau2020MNRAS.tmp..448C} and \cite{Konno2018_SR4}. Fixing redshifts at $z=$1.47/1.19 and $z=$0.84/0.63 gives [O\textsc{ii}] and [O\textsc{iii}] line luminosities that are also in good agreement with \cite{Khostovan2016MNRAS.463.2363K} and \cite{Hayashi2018PASJ...70S..17H}, as described in Sect. \ref{subsec:emission_line_radio}. 

The EW$_{0}$ and SFR values derived for all three redshifts lie within the range of reported values for other samples in the literature. However, these physical parameters vary substantially both in this work and the literature, which is likely due to a wide range of galaxy types and survey selection functions (see e.g. \citealt{khostovan2019MNRAS.489..555K} and \citealt{calhau2020MNRAS.tmp..448C}). If we assume the candidates to be low-$z$ interlopers, we are again studying a subset of sources of the [O\textsc{ii}] and [O\textsc{iii}] emitter population, which is likely biased, and thus these sources do not necessarily have the same physical properties. We therefore refrain from making any further comparisons between the obtained EW$_{0}$ and SFRs in an attempt to support the evidence for the low-$z$ nature of a high fraction of these sources. Remarkable, however, are the stellar masses for fixed $z=6.57$, which are extremely high (M$_*$ > 10$^{11}$ M$_{\odot}$), especially for the sources in the NB921 catalogue. The majority of these sources are therefore unlikely to be LAEs. 

The derived source properties are also subject to systematic effects from the SED fitting models used and the possible biases introduced by the sample selection. \textsc{Bagpipes} is constrained by the set input parameters, prior distributions, and model dependencies. Despite the limitations, the models can often successfully recreate the observed SEDs. However, spectroscopic observations are necessary to confirm the redshifts and derived physical properties.

\subsection{Radio and IR-optical stacking}
\label{subsec:stacking}

\begin{figure}
\centering
   \includegraphics[width=\columnwidth, trim={0.0cm 0.0cm 0.0cm 0.0cm}, clip]{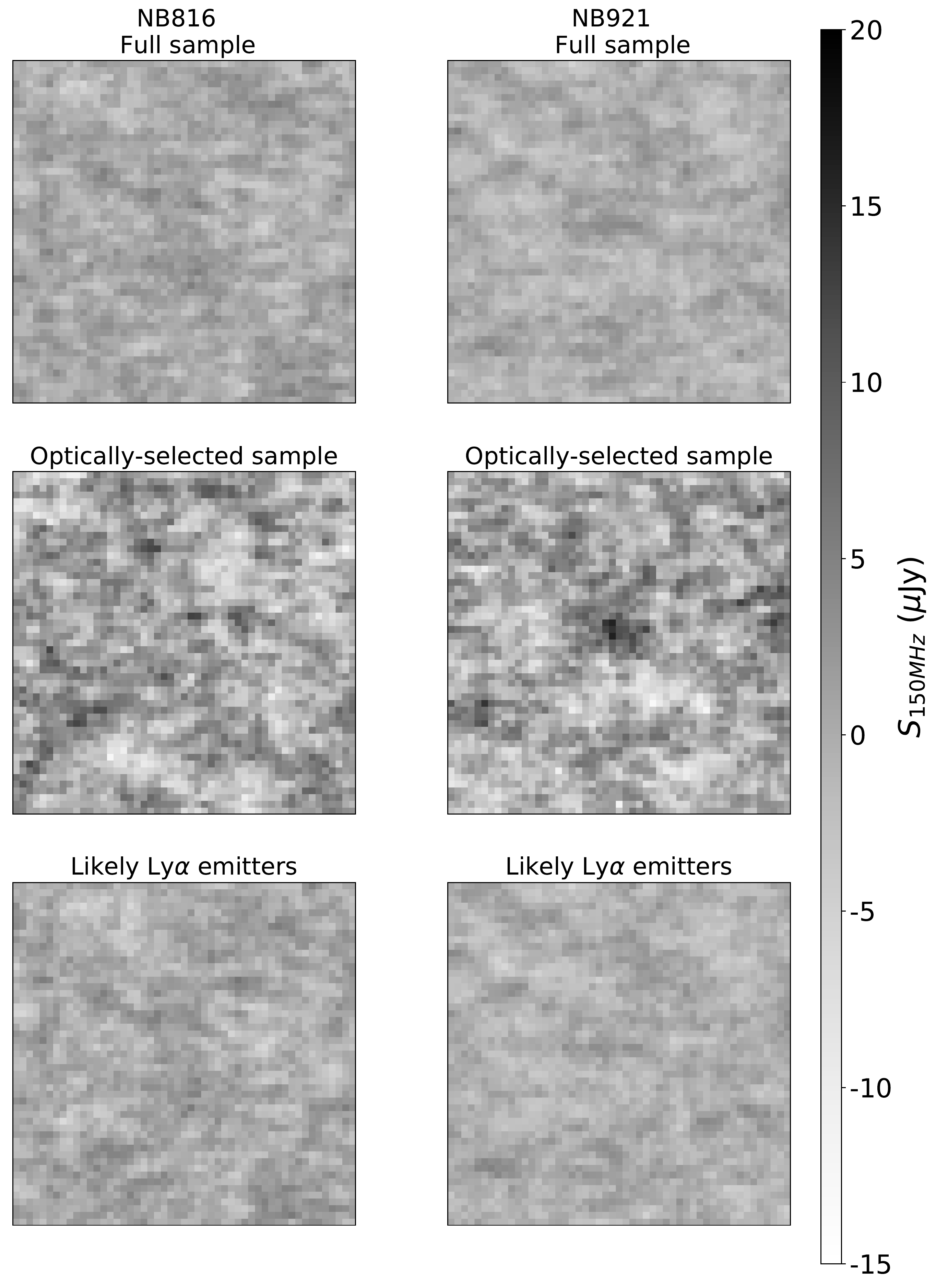}
     \caption{Stacked LOFAR cutouts of LAE candidate samples. Left: NB816 full sample (226 sources; upper left), optically selected sample (53 sources; middle left), and likely Ly$\alpha$ emitters (177 sources; bottom left). Right: NB921 full sample (322 sources; upper right), optically selected sample (49 sources; middle right), and likely Ly$\alpha$ emitters (279 sources; bottom right). The LOFAR-detected sources and sources not overlapping with LOFAR coverage have been removed.} 
     \label{fig:radio_stacks}

\end{figure}

To investigate the radio properties of the LAE candidates not detected by LOFAR, first we stacked the LOFAR images to set observational constraints on the average radio properties of the samples. SILVERRUSH sources outside the area covered by the multi-wavelength LoTSS data (see Sect. \ref{subsec:lofar_optical_phot}) are removed from the catalogue (22 and three sources for the NB921 and NB816 catalogues, respectively). The five sources in the LOFAR-detected sample in the NB921 catalogue are also removed from the stacking data set (see Sect. \ref{sec:radio_silverrush}). For both NB816 and NB921, first the full sample was stacked (containing 226 and 322 sources, respectively) and then the optically selected samples (containing 53 and 49 sources) were stacked separately; these results are shown in the top and middle panels of Fig. \ref{fig:radio_stacks}. We assumed the sources to be unresolved in the 6" resolution radio images, and could therefore obtain the median flux from the peak pixels. The average rms noise level for individual radio images is $\sim$25 $\mu$Jy beam$^{-1}$. The stacked radio flux density, rms, and resulting S/N obtained in each sample are summarised in Table \ref{tab:stacking}. None of the stacked radio flux densities are significant (for all S/N < 2). Therefore, we can place a 3$\sigma$ upper limit on the radio flux density of $\sim$4.8 $\mu$Jy for the sources in the full sample and $\sim$12.0 $\mu$Jy for sources in the optically selected sample. 

Furthermore, we stacked the four and five optically selected sources from NB816 and NB921, respectively, that seemed to be likely LAEs from SED fitting together with the 173 and 274 sources in the non-matched SILVERRUSH samples, under the assumption that they are LAEs. The non-matched sources are fainter and bluer and therefore less likely to be low-redshift contaminants (see Fig. \ref{fig:magnitude_compare} and Sect. \ref{subsec:lae_selection_discussion}). The stacks again yield no detection with a limiting radio flux density of 5.7 $\mu$Jy and 4.8 $\mu$Jy (3$\sigma$) for NB816 and NB921, respectively (see the bottom panels of Fig.~\ref{fig:radio_stacks}). When assuming their redshifts to be $z=$5.7 and 6.6, the radio flux density can be converted to 2$\sigma$ radio luminosity upper limits of 8.2$\times$10$^{23}$ and 8.7$\times$10$^{23}$ W Hz$^{-1}$, respectively. Using the low-$z$ L$_{150\rm{MHz}}$-SFR relation from \cite{gurkan2018MNRAS.475.3010G}, an estimated 2$\sigma$ upper limit on the SFR can be placed on these LAEs. Our derived upper limits are $\sim$53 and 56 M$_{\odot}$ yr$^{-1}$ on the $z=5.7$ and $z=6.6$ LAEs, respectively. The currently known relations between L$_{150\rm{MHz}}$ and the SFR are derived from low-$z$ galaxies; therefore, this is our current best SFR estimate. However, the work of \cite{Smith2020} suggests that there is no strong redshift evolution. Similar SFRs are obtained with the L$_{150\rm{MHz}}$-SFR relations from \cite{wang2019A&A...631A.109W}, with differences of a few solar masses per year. This upper limit is higher than the expected SFR of LAEs (see e.g. \citealt{calhau2020MNRAS.tmp..448C}), so, to be able to place stronger constraints on the SFR, a larger sample of LAEs, or deeper observations, are necessary.

Next, we stacked the optical bands, ranging from \textit{u} to \textit{y}, and find no detection in the stacked full sample, with SpARCS \textit{u}, HSC \textit{g}, and HSC \textit{r} band 3$\sigma$ limiting magnitudes of $\sim$27.6, 25.1, and 24.5, respectively (using a 3" aperture), indicating that this population of likely $z\sim$6 SILVERRUSH galaxies without radio detection is consistent with $z=6.6$ down to this magnitude limit, as expected from the initial selection method. These non-detections, however, do not rule out the possibility of them being low-z interlopers. We find similar results when we stack optical images for the optically selected sample, with measurements of \textit{u, g,} and \textit{r} down to 3$\sigma$ limiting magnitudes of 26.8, 24.2, and 23.7, respectively. In contrast, SED fitting on the full panchromatic photometry in the optically selected sample suggested a large contamination rate (see Sect. \ref{subsec:sed_fit_full_pop}). These sources in the optically selected sample are typically faint in the optical and are primarily detected in the $\chi^2$ image because of their high IR flux, indicating that IR observations are critical for identifying low-redshift interlopers.

Finally, we investigated the NIR and mid-IR data for the full LAE candidate sample (322 and 226 sources in NB921 and NB816, respectively) by stacking the J and K images (1.25 and 2.21 $\mu$m) obtained by UKIDSS, as well as the mid-IR (3.6, 4.5, 5.8, and 8.0 $\mu$m) images obtained by Spitzer-SWIRE. We find a detection (S/N > 3) in all, except for the 5.8 $\mu$m channel, with only a weak detection (S/N$\sim$3) in the 8 $\mu$m channel; this, again, indicates an absence of strong AGN contribution.

\begin{table}
\caption{Number of sources, radio flux density, rms, and resulting S/N for LOFAR stacked images of full and optically selected sample in NB816 and NB921.} 
\label{tab:stacking}      
\centering  
\resizebox{\columnwidth}{!}{
\begin{tabular}{ c | c | c | c | c }  
\hline \hline
Sample & Num. sources & S$_{150 MHz}$ ($\mu$Jy) & rms ($\mu$Jy) & S/N \\
\hline
\multicolumn{5}{l}{\textit{NB816}} \\
\hline
Full sample & 226 & 2.4 & 1.7 & 1.4 \\
Optically selected sample & 53 & $-$2.6 & 4.0 & - \\
Likely LAE sample & 177 & 0.02 & 1.9 & 0.01 \\
\hline
\multicolumn{5}{l}{\textit{NB921}} \\
\hline
Full sample & 322 & 1.8 & 1.5 & 1.2 \\
Optically selected sample & 49 & 7.4 & 3.9 & 1.9 \\
Likely LAE sample & 279 & 0.2 & 1.6 & 0.1 \\
\hline \hline
\end{tabular}
} 
\end{table}

\section{Discussion}
\label{sec:discussion}

In this section, we discuss the LOFAR detection rate (Sect. \ref{subsec:lofar_detection_rate}), the LAE sample selection, and possible biases of and improvements to the selection process (Sect. \ref{subsec:lae_selection_discussion}).

\subsection{LOFAR detection rate of SILVERRUSH LAEs}
\label{subsec:lofar_detection_rate}

No $z > 6$ radio galaxy candidates have been found in the cross-matched LoTSS and SILVERRUSH survey. In this section, we discuss this in light of predictions made for the number densities of such sources. The expected number of radio galaxies in LoTSS in the redshift slices 5.72$\pm$0.05 and 6.57$\pm$0.05, from models presented by \cite{Saxena2017MNRAS.469.4083S}, are 5.3$\pm$0.4 and 1.5$\pm$0.1 sources per deg$^2$, respectively. From these models, we would expect about nine radio galaxies with 6.52 < z < 6.62 to be detected in the $\sim$6 deg$^2$ overlap region between SILVERRUSH and LoTSS. However, the LAE fraction in brighter galaxies at redshifts 6 < z < 8 is shown to be only 10-20\% in the study by \cite{Schenker2012ApJ...744..179S}, and the distribution of Ly$\alpha$ luminosities in high-$z$ radio galaxies ($z > 3$) is poorly constrained due to small sample sizes (see e.g. \citealt{Saxena2019MNRAS.489.5053S}). Assuming this LAE fraction, only $\sim$1 out of the nine predicted sources is expected to be selected by SILVERRUSH and detected by LOFAR. Given this expectation of only one detection, it is not surprising that we do not find any radio emitting LAEs at $z\sim6$. Furthermore, there are still many uncertainties in the expected number of radio detections, especially at high redshift. The LoTSS Deep Field observations probe the faint end of the radio luminosity function (RLF) due to the small volume and high sensitivity, which is poorly constrained by the models (see \citealt{Saxena2017MNRAS.469.4083S}). Therefore, the uncertainties in the RLF are critical for the number density predictions in this paper. These points together provide a plausible explanation for the absence of $z > 6$ radio galaxies in the studied sample. Based on these numbers, a larger area of overlapping coverage would be necessary to find many $z > 6$ radio galaxies.

\subsection{LAE selection methods}
\label{subsec:lae_selection_discussion}

\begin{figure}
\centering
   \includegraphics[width=\columnwidth, trim={0.0cm 0cm 0cm -1.0cm}, clip]{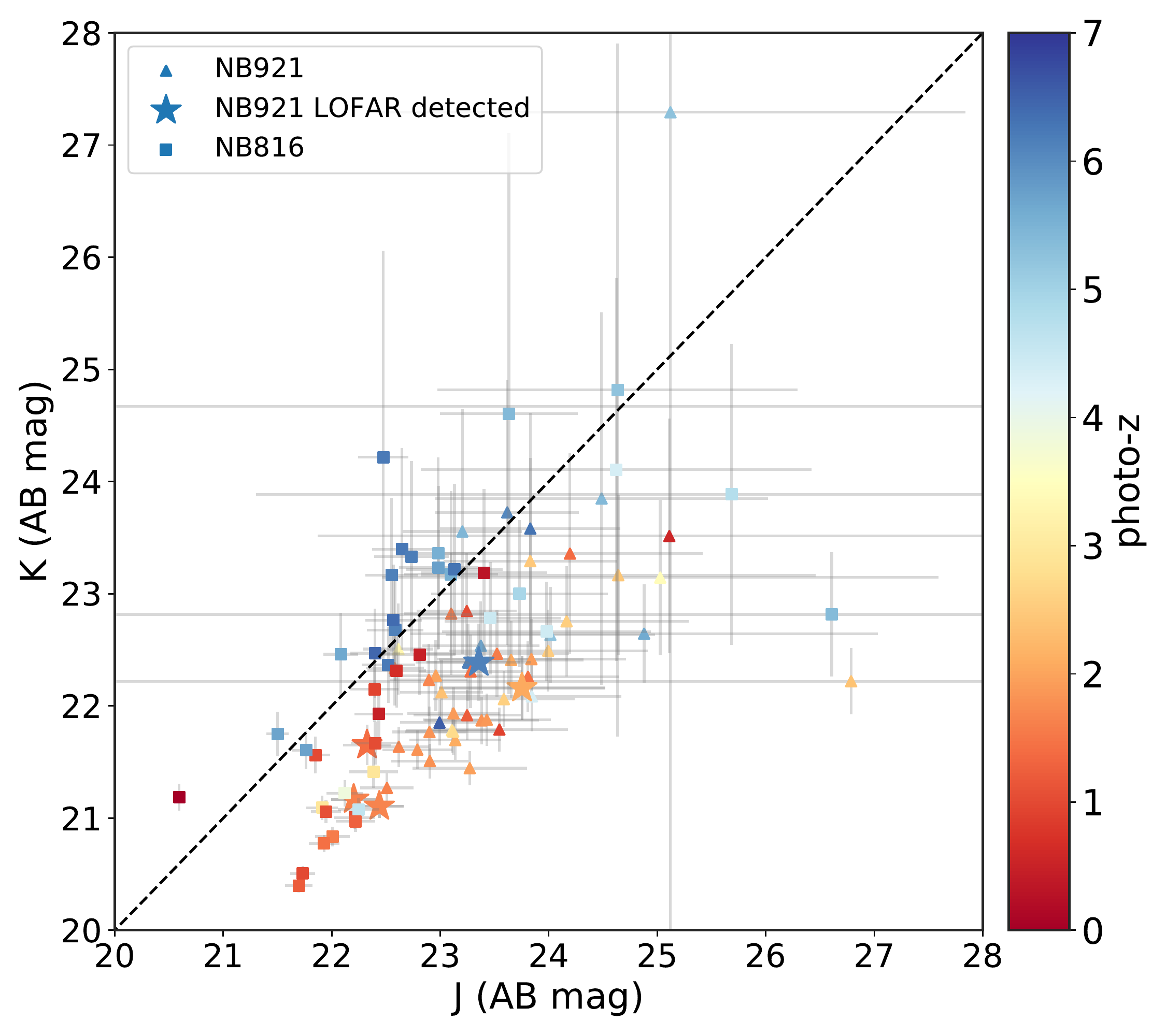}
     \caption{UKIDSS J and K band magnitude of the optically selected sample and LOFAR-detected sample, with colour-coded photo-$z$s from \citetalias{Duncan2020}. The diagonal (dashed) line indicates the J$-$K $\leq$ 0 criterion. The sources above this line satisfy this criterion and also have notably higher photo-z values.}
     \label{fig:J_K_mag}
\end{figure}

It is possible that, with the optical and IR selected samples, we are probing galaxies that are more likely to be low-redshift interlopers than the SILVERRUSH samples that do not appear in our optical and IR catalogue. Very dusty low-redshift galaxies could, for instance, have a Balmer break at 4000 $\r{A},$ mimicking the Lyman break (see e.g. \citealt{matthee2014MNRAS.440.2375M, Matthee2017MNRAS.472..772M}). These galaxies often have red J$-$K colours, so constraining J$-$K to be flat or blue (J $-$ K $\leq$ 0) prevents contamination from these dusty low-redshift interlopers. The J$-$K colours of the optically selected samples and the LOFAR-detected sample are shown in Fig. \ref{fig:J_K_mag}. For both the NB816 and NB921 optically selected samples, eight sources (out of the 53 and 58 sources, respectively) do not have J or K magnitudes. Only three sources in NB921 satisfy J$-$K $\leq$ 0, and these three all have high photo-$z$ values (see Fig. \ref{fig:J_K_mag}). No sources within the LOFAR-detected sample satisfy this criterion. The effect is even more apparent for NB816, where the diagonal line divides the low photo-$z$ candidates from the high photo-$z$ ones, except for one extremely low-$z$ source. Therefore, NIR observations are of great value in identifying the low-$z$ interlopers. However, we note that a strict J$-$K $\leq$ 0 cut could cause red LAEs to be rejected from the sample. 

\cite{Shibuya2018_SRIII} have shown that the contamination rate depends on the NB magnitude, where populations with lower magnitudes have higher contamination rates. The sources matched with our optical catalogue are brighter and likely redder on average than the full LAE sample due to the shallower HSC observations and the source extraction from the deep $\chi^2$ image in the optical catalogue. The source extraction is based on significant detection in the $\chi^2$ stack of either optical and NIR or IRAC Channel 1 and 2 images. The SILVERRUSH sources that are missing from our optical sample will therefore be the ones that are fainter in the optical bands and not red enough to be detected in the NIR or mid-IR. The $\chi^2$ detection image did not include any \textit{y}-band data; this could significantly contribute  to whether or not the real LAEs are detected since, at $6<z<7,$ sources are not detected blue-wards of the \textit{y} band where the Lyman break is situated. Therefore, the non-matched sources are more likely LAEs. If we assume that all the non-matched sources are true LAEs, then we obtain contamination rate lower limits of 23 and 20 \% for the full NB816 and NB921 catalogues, respectively, which would agree with the quoted 30\% contamination rate from the spectroscopic follow-up by SILVERRUSH (see \citealt{Shibuya2018_SRIII}). 

\begin{figure*}
    \centering
    \begin{subfigure}{\columnwidth}
        \centering
        \textbf{Stack of likely Ly$\alpha$ emitters}\par\medskip
        \vspace{-0.2cm}
        \includegraphics[width=\textwidth, trim={0cm 0cm 0cm 0.0cm}, clip]{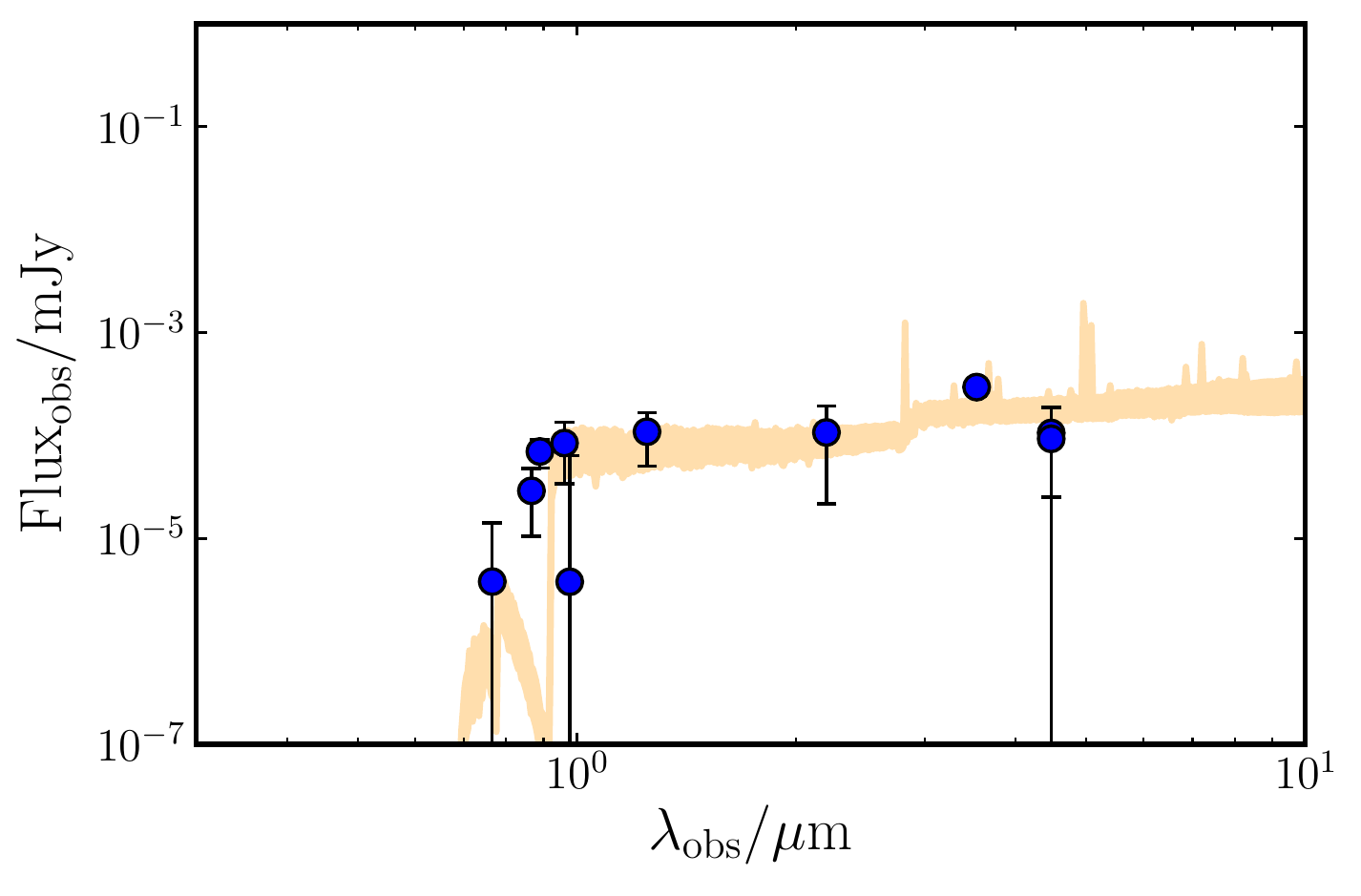}
    \end{subfigure}
     ~ 
    \begin{subfigure}{\columnwidth}
        \centering
        \textbf{Stack of likely contaminants}\par\medskip
        \vspace{-0.2cm}
        \includegraphics[width=\textwidth, trim={0cm 0cm 0cm 0.0cm}, clip]{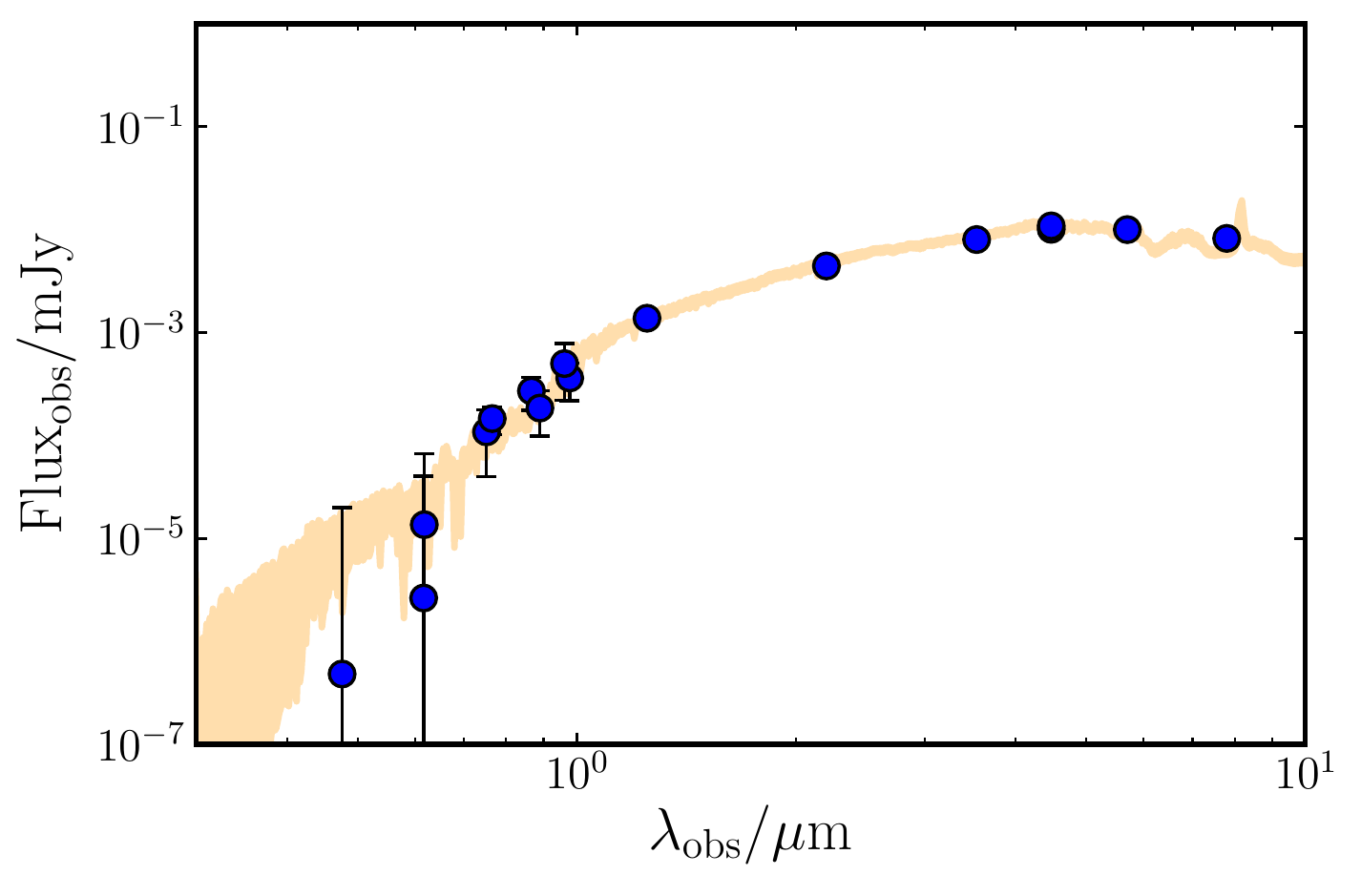}
    \end{subfigure}
    \caption{\label{fig:stacked_seds} Stacked SEDs (blue) with posterior median SED models (yellow). Left: Stacked likely $z=6.6$ LAE candidates in the NB921 catalogue, consisting of the non-matched sample and probable LAEs in the matched sample. Right: Stacked likely contaminants in the NB921 matched optical sample, fitted with fixed $z=1.47$. The stack of likely contaminants shows brighter NIR and mid-IR fluxes, highlighting the importance of using NIR data in LAE selection.}
\end{figure*}

To highlight the difference between the likely LAEs and the contaminants in the NB921 catalogue, we show the stacked SEDs in Fig. \ref{fig:stacked_seds}. The SED of the stacked assumed LAE sample is consistent with $z=6.6$, whereas the SED of the contaminant sample is consistent with $z=1.47$. For LAEs at $z=6.6$, the IRAC 3.6 $\mu$m filter corresponds to the [O\textsc{iii}]+H$\beta$ emission line (see \citealt{Smit2015ApJ...801..122S}).  Since the high EWs of the Ly$\alpha$ and [O\textsc{iii}]+H$\beta$ lines might not be represented sufficiently well by the nebular emission line models in \textsc{Bagpipes}, the NB921 filter and IRAC 3.6 $\mu$m filter have not been included in the fits. The J$-$K colours are -0.02$\pm$1.05 and 1.27$\pm$0.11 for the stacked assumed LAEs and contaminants, respectively. This again highlights the advantage of selecting galaxies for which the NIR slope satisfies J $-$ K $\leq$ 0 when searching for high-redshift LAEs. From the \textsc{Bagpipes} models, we obtained stellar masses of 10$^{9.6 \pm 0.2}$ and 10$^{10.5 \pm 0.1}$ M$_{\odot}$ for the SED fit of stacked LAEs and stacked contaminants respectively, and SFRs of 8.5$^{+5.5}_{-3.1}$ and $<0.4$ M$_{\odot}$ yr$^{-1}$, where the upper limit is the 84th percentile of the SFR posterior. This is consistent with the upper limit on the LAE SFR obtained from the corresponding radio stacks (see Sect. \ref{subsec:stacking}).

As we concluded in Sect. \ref{subsec:sed_fit_full_pop}, the majority of the LAE candidates that have a match in our optical catalogues are likely to be low-$z$ interlopers of [O\textsc{ii}] and [O\textsc{iii}] emitters, or possibly H$\alpha$ at $z=0.40$. If the sample is indeed contaminated by a large fraction of lower-redshift line emitters, other NB studies might have also been affected by large fractions of contaminants, possibly leading to a misrepresentation of the physical properties of LAEs.

To minimise the effects of contamination, one option for future studies is to consider the optical colour selections used to select LAE candidates. In the SILVERRUSH studies, a forced-aperture catalogue is used for statistical studies (see \citealt{Shibuya2018}). This catalogue is generated with apertures at fixed positions, with a typical 3$\sigma$ detection requirement that is lowered to 2$\sigma$ for the g and r bands. Colour selections are then imposed to select LAEs. The selections made are z$-$NB921>1.8 for $z=6.6$ LAEs, which corresponds to EW$_{0,\text{Ly}\alpha}$  > 14 $\r{A}$ for the z$\sim$6.6 sample. We now compare the numbers that would be derived from the forced SILVERRUSH catalogue, as opposed to the combined forced and unforced catalogue used in the rest of this study. The forced SILVERRUSH catalogue contains 48 sources in NB921 and 130 in NB816. The number of sources in the forced catalogues compared to the unforced catalogue is substantially lower; therefore only two LOFAR-detected sources are found in NB921, and no sources are found in NB816. These two matched sources are ILTJ160227.18+544759.8 and ILTJ160930.42+544435.9, which are also likely $z=1.5$ [O\textsc{ii}] emitters (see the top two panels of Fig. \ref{fig:sed_3_sources}). From the optical sample, there are three and 11 in the forced catalogues of NB921 and NB816, respectively. All three sources in NB921 have photo-$z$s of $\sim$1.7$\pm$0.7, suggesting that these are low-redshift interlopers. However, of the 11 sources in the NB816 forced catalogue, ten have a photo-$z$ in the range 4.5-5.8$\pm$1.5 (the one remaining source has a photo-$z$ of 0.5$\pm$0.9), suggesting that the LAE fraction is likely high. This implies that, at least in the case of $z=5.7$ LAEs, the contamination rate could be significantly lowered by making stricter cuts. 

Many LAE NB surveys apply similar cuts for selection, but differ in their strictness on the cuts and the availability of multiple bands (e.g. \citealt{Ouchi2008ApJS..176..301O, Santos2016MNRAS.463.1678S, Matthee2015MNRAS.451..400M}).  Other studies have also shown that the majority of LAEs have EW$_0$ > 50 $\r{A}$ (e.g. \citealt{Ouchi2008ApJS..176..301O}). However, \cite{Sobral2018MNRAS.476.4725S} also show, by using a large compilation of spectroscopic redshifts, that, even though their sample of high EW candidate line-emitters is dominated by likely LAEs, there is still a significant population of H$\alpha$, [O\textsc{iii}]+H$\beta$, and [O\textsc{ii}] emitters. Since the observed EW$_0$ distribution in these lines is similar to Ly$\alpha$ (see e.g. \citealt{Khostovan2016MNRAS.463.2363K, Hayashi2018PASJ...70S..17H}), it can be challenging to clean the samples further using just NB and BB optical surveys. The contamination rates we obtain are comparable to those found by \cite{matthee2014MNRAS.440.2375M}, who find a 90\% contamination in LAE samples selected on the Lyman break criteria. However, their survey targeted higher redshifts of $z > 7$, so contamination is expected to be higher.

This work further highlights the fact that ancillary IR observations are critical for the LAE sample selection. Although the J to 4.5 $\mu$m data in ELAIS-N1 are significantly shallower than those available in smaller fields, such as COSMOS, they can still play a important role in minimising interlopers in wide field LAE samples. Decreasing the contamination rate is critical for improving the constraints on the Ly$\alpha$ LF and clustering measurements, as well as for reducing the time and expense required for spectroscopic follow-up observations. 

\section{Summary}
\label{sec:summary}

In this paper, we have studied the LOFAR and optical and IR properties of SILVERRUSH LAEs in ELAIS-N1 using the latest LoTSS Deep Fields observations and the cross-matched multi-wavelength catalogue that ranges from UV to FIR. The publicly available SILVERRUSH catalogues of ELAIS-N1 consist of 229 and 349 LAE candidates at $z=5.7$ and $z=6.6$, respectively, identified with the HSC NB816 and NB921 narrowbands. We found five LOFAR-detected LAEs in the NB921 SILVERRUSH catalogue. These five sources have LOFAR radio fluxes in the range of 0.09-0.35 mJy. We performed SED fitting on the radio-detected sources using \textsc{Bagpipes} and find that all five sources are likely $z=1.5$ [O\textsc{ii}] emitters, based on $\chi^2$ analysis and derived physical properties. We therefore do not find any promising $z > 6$ radio galaxy candidates and cannot put any constraints on the AGN fraction of LAE at high redshift. 

In light of the high contamination rate of radio-selected SILVERRUSH sources, we performed the same SED fitting procedure on the wider LAE population by cross-matching the SILVERRUSH NB816 and NB921 catalogues with the LoTSS Deep Field multi-wavelength catalogue. We find 53 and 58 sources with matches in the LoTSS multi-wavelength catalogue, out of the 229 and 349 SILVERRUSH LAE candidates, respectively. The resulting $\Delta \chi^2$ values suggest very high contamination rates of 81\% (92\%) and 81\%\ (91\%)\ for the NB816 and NB921 catalogue sample, respectively, where the values in the parentheses include the sources that are also likely contaminants but not statistically significant. This is consistent with the photo-$z$s from \citetalias{Duncan2020} given the large photo-z uncertainties at these high redshifts and faint magnitudes. Of all the contaminated sources, 63$\pm$31\% are likely to be [O\textsc{ii}] and 6$\pm$31\% are likely to be [O\textsc{iii}] emitters for NB816, and similarly 55$\pm$28\% are likely to be [O\textsc{ii}] and 17$\pm$28\% [O\textsc{iii}] emitters for NB921 (with the remainder inconclusive). However, the non-matched sources are fainter and bluer and therefore more likely to be true LAEs, which would be in line with the 30\% contamination rate obtained from spectroscopic follow up by SILVERRUSH \citep{Shibuya2018_SRIII}.

Most importantly, in agreement with \cite{matthee2014MNRAS.440.2375M, Matthee2017MNRAS.472..772M}, we find that a J$-$K $\leq$ 0 cut can identify a large fraction of red and dusty low-$z$ interlopers, which are not removed by the standard LAE criteria due to the low optical S/Ns and the Balmer break mimicking the Lyman break. These results highlight the need for ancillary IR observations in LAE sample selection to minimise the amount of interlopers.

We stacked radio image cutouts in the positions of all LAE candidates and find no significant detection with LOFAR when removing the radio-detected sources. Removing the likely contaminating sources and stacking the LOFAR data for the four and five remaining matched sources in NB816 and NB921 together with the 173 and 274 non-matched sources yields a 2$\sigma$ upper limit on the SFR of $\sim$53 and 56 M$_{\odot}$ yr$^{-1}$, respectively. In the optical stacked \textit{u, g}, and \textit{r} measurements, we do not find a detection down to $\sim$27.6, 25.1, and 24.5 magnitudes, respectively, which is consistent with the Lyman break at $z=6.6$. The stacked IRAC SWIRE 3.6 and 4.5 $\mu$m and UKIDSS J and K images do yield a detection (S/N>3), but there is no significant detection (S/N$\sim$2) in the IRAC SWIRE 5.8 and 8.0 $\mu$m measurements. 

The current 180 hours of observations of ELAIS-N1 are expected to reach 500 hours within the next two years. This will result in factor of $\sim$2 deeper observations down to 11 $\mu$Jy beam$^{-1}$, enabling the study of the fainter and higher-redshift radio population. Furthermore, the WEAVE-LOFAR spectroscopic survey (\citealt{smith2016sf2a.conf..271S}; commissioned for early 2021) will obtain around 10$^6$ optical spectra of LOFAR-detected radio sources, allowing for an accurate redshift determination and source classification. This survey will provide the opportunity to reveal high-redshift ($z > 6$) radio galaxies, which are necessary for advancing our current understanding of galaxy formation and evolution and can be used as probes of the EoR. Follow-up observations with existing 10m class optical telescopes and upcoming facilities, such as the \textit{James Webb Space Telescope} and the \textit{European Extremely Large Telescope}, will enable detailed studies of these high-redshift galaxies for the first time.

\begin{acknowledgements}

{KJD and HJAR acknowledge support from the ERC Advanced Investigator programme NewClusters 321271. JS and PNB are grateful for support from the UK STFC via grant ST/R000972/1. IP and Marco Bondi acknowledges support from INAF under PRIN SKA/CTA FORECaST and MAIN STREAM SAUROS projects. K.M. has been supported by the National Science Centre (UMO-2018/30/E/ST9/00082). Matteo Bonato acknowledges support from the Ministero degli Affari Esteri della Cooperazione Internazionale - Direzione Generale per la Promozione del Sistema Paese Progetto di Grande Rilevanza ZA18GR02. RK acknowledges support from the Science and Technology Facilities Council (STFC) through an STFC studentship via grant ST/R504737/1.

This paper is based (in part) on data obtained with the International LOFAR Telescope (ILT) under project codes LC0 015, LC2 024, LC2 038, LC3 008, LC4 008, LC4 034 and LT10 01. LOFAR \citep{vanHaarlem2013A&A...556A...2V} is the Low Frequency Array designed and constructed by ASTRON. It has observing, data processing, and data storage facilities in several countries, which are owned by various parties (each with their own funding sources), and which are collectively operated by the ILT foundation under a joint scientific policy. The ILT resources have benefitted from the following recent major funding sources: CNRS-INSU, Observatoire de Paris and Universit\'e d'Orl\'eans, France; BMBF, MIWF-NRW, MPG, Germany; Science Foundation Ireland (SFI), Department of Business, Enterprise and Innovation (DBEI), Ireland; NWO, The Netherlands; The Science and Technology Facilities Council, UK; Ministry of Science and Higher Education, Poland.

The Hyper Suprime-Cam (HSC) collaboration includes the astronomical communities of Japan and Taiwan, and Princeton University. The HSC instrumentation and software were developed by the National Astronomical Observatory of Japan (NAOJ), the Kavli Institute for the Physics and Mathematics of the Universe (Kavli IPMU), the University of Tokyo, the High Energy Accelerator Research Organisation (KEK), the Academia Sinica Institute for Astronomy and Astrophysics in Taiwan (ASIAA), and Princeton University. Funding was contributed by the FIRST program from Japanese Cabinet Office, the Ministry of Education, Culture, Sports, Science and Technology (MEXT), the Japan Society for the Promotion of Science (JSPS), Japan Science and Technology Agency (JST), the Toray Science Foundation, NAOJ, Kavli IPMU, KEK, ASIAA, and Princeton University.}

\end{acknowledgements}

\bibliographystyle{aa}
\bibliography{bibliography.bib}

\begin{appendix}

\section{\textsc{Bagpipes} model parameters}
Table \ref{tab:bagpipes_parameters} shows the parameters we used to fit to our data in \textsc{Bagpipes} \citep{Carnall2018}. The SFH is modelled using a double-power-law given by

\begin{equation}
    SFR(t) \propto \Big[\Big(\frac{t}{\tau}\Big)^{\alpha}+\Big(\frac{t}{\tau}\Big)^{-\beta}\Big]^{-1}
,\end{equation}

where $\alpha$ is the falling power slope, $\beta$ is the rising power slope , and $\tau$ is the time of peak star formation. A prior distribution uniform in log$_{10}$ is used for $\alpha$ and $\beta$. For further information on the model components visit \footnote{\url{https://bagpipes.readthedocs.io/en/latest/model_components.html}}.

\begin{table}
\caption{Overview of parameters used in this work for fitting to our data in \textsc{Bagpipes}.}  
\label{tab:bagpipes_parameters}      
\centering  
\begin{tabular}{ c | c | c}  
\hline \hline
\textbf{Symbol} & \textbf{Prior value range} & \textbf{Parameter description} \\
\hline
\multicolumn{3}{l}{\textit{SFH}} \\
\hline
$\tau$ & [0,15] Gyr & peak time of SFR \\
$\alpha$ & [0.01,1000] & rising power slope \\
$\beta$ & [0.01,1000] & falling power slope \\
M$_{*,\text{formed}}$ & [7,14] log$_{10}$(M$_{\odot}$) &  \multicolumn{1}{p{4cm}}{\centering {amount of stellar mass} \\ {formed} }  \\
metallicity & [0,2.5] M$_{\odot}$  & metallicity of galaxy \\
\hline
\multicolumn{3}{l}{\textit{Dust}} \\
\hline
type & Calzetti &  attenuation law \\
Av & [0,6] & \multicolumn{1}{p{4cm}}{\centering {absolute attenuation} \\ {in V band} }  \\
umin & [0.1,25] & \multicolumn{1}{p{4cm}}{\centering {lower limit of starlight} \\ {intensity distribution} }  \\
$\gamma$ & [0.0001,1.0] & fraction of stars at umin\\
qpah & [0.1,4.5] & PAH mass fraction \\
\hline
\multicolumn{3}{l}{\textit{Nebular}} \\
\hline
log$_{10}$(U) & -3 & ionisation parameter \\
\hline \hline
\end{tabular}
\end{table}

\section{SED fits of LOFAR-detected sources}
\begin{figure*}
\centering
   \includegraphics[width=0.9\textwidth, trim={0cm 2.5cm 0.5cm 0.5cm}, clip]{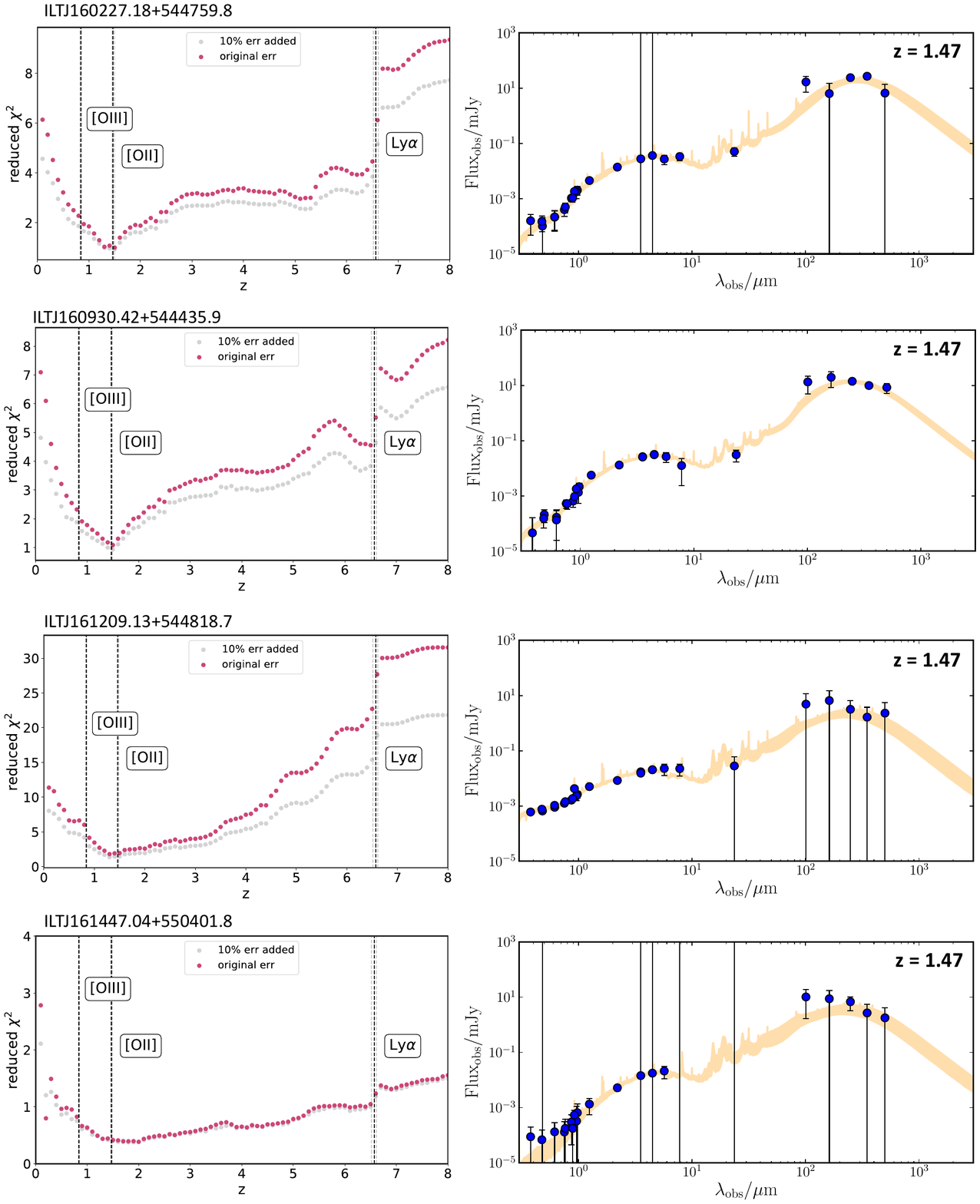}
     \caption{SED fitting results obtained using \textsc{Bagpipes} for ILTJ160227.18+544759.8, ILTJ160930.42+544435.9, ILTJ16209.13+544818.7, and ILTJ161447.04+550401.8 (in order from top to bottom). Left: Reduced $\chi^2$ as a function of redshift. The redshifts corresponding to the [O\textsc{iii}], [O\textsc{ii}], and Ly$\alpha$ emission lines are indicated, and both resulting reduced $\chi^2$ values using normal error (pink) and 10\% extra flux error (grey) are shown. Right: Posterior median model obtained from \textsc{Bagpipes} (yellow) with $z=1.47$ and over-plotted photometric fluxes as  a function of wavelength (blue).}
     \label{fig:sed_3_sources}
\end{figure*}

The SED fits and reduced $\chi^2$ as a function of redshift are given in Fig. \ref{fig:sed_3_sources} for four sources in the LOFAR-detected sample.

\end{appendix}

\end{document}